%% file: master.tex
\documentclass[aps,prb,superscriptaddress,tightenlines,nofootinbib,floatfix,longbibliography,notitlepage]{revtex4-2}
\usepackage[left=18mm,right=19mm,top=23mm,bottom=16mm]{geometry}
\usepackage{amsmath,amssymb}
\usepackage{bm}
\usepackage{bbold}
\usepackage{comment}
\usepackage{graphicx}
\usepackage{siunitx}
\usepackage[dvipsnames]{xcolor}
\usepackage{slashed}
\usepackage{physics}
\usepackage[
    colorlinks=true,
    allcolors=blue
]{hyperref}
\usepackage{cleveref}
\usepackage{tikz}
\usetikzlibrary{arrows,matrix,arrows.meta}
\tikzstyle{arrow} = [very thick,->,>=stealth]
\usepackage{adjustbox}
\usepackage[justification=RaggedRight]{subcaption}
\providecommand{\repositoryInformationSetup}{} 
\repositoryInformationSetup

\include{macros}

\input{affiliations}

\begin{document}
\title{Fermionic Sign Problem Minimization by Constant Path Integral Contour Shifts}
\author{Christoph G\"antgen} \affiliation{\ias}\affiliation{\casa}\affiliation{\bonn}
\author{Evan Berkowitz} \affiliation{\ias}\affiliation{\casa}\affiliation{\jsc}
\author{Thomas Luu}  \affiliation{\ias} \affiliation{\bonn} \affiliation{\ikp}
\author{Johann Ostmeyer} \affiliation{\liverpool}
\author{Marcel Rodekamp} \affiliation{\ias}\affiliation{\casa}\affiliation{\bonn}\affiliation{\jsc} 

\date{\today}

\begin{abstract}
	The path integral formulation of quantum mechanical problems including fermions is often affected by a severe numerical sign problem.
	We show how such a sign problem can be alleviated by a judiciously chosen constant imaginary offset to the path integral. Such integration contour deformations introduce no additional computational cost to
	the Hybrid Monte Carlo algorithm, while its effective sample size is greatly increased. This makes otherwise unviable simulations efficient for a wide range of parameters.
	Applying our method to the Hubbard model, we find that the sign problem is significantly reduced. 
    Furthermore, we prove that it vanishes completely for large chemical potentials, a regime where the sign problem is expected to be particularly severe without imaginary offsets.
	In addition to a numerical analysis of such optimized contour shifts, we analytically compute the shifts corresponding to the leading and next-to-leading order corrections to the action. We find that such simple approximations, free of significant computational cost, suffice in many cases.
\end{abstract}
\maketitle
\input{section/introduction}

\input{section/theory}

\input{section/method}
\input{section/results}
\input{section/conclusion}

\begin{acknowledgments}
	We thank Neill Warrington for many helpful discussions related to this work
	as well as Timo Lähde for his valuable comments.
	This work was funded in part by the
	Deutsche Forschungsgemeinschaft (DFG, German Research Foundation) through the funds provided to the
	Sino-German Collaborative Research Center “Symmetries and the Emergence of Structure in QCD” (NSFC
	Grant No. 12070131001, DFG Project-ID 196253076
	– TRR110) as well as the STFC Consolidated Grant
	ST/T000988/1.
	We gratefully acknowledge the computing time
	granted by the JARA Vergabegremium and provided on
	the JARA Partition part of the supercomputer JURECA
	at Forschungszentrum Jülich.
\end{acknowledgments}
\appendix

\input{section/appendix}

\bibliography{references}

\end{document}

%% file: macros.tex
\usepackage{xspace}
\usepackage{bbm}

\definecolor{fzjblue1}{cmyk}{1.00,0.40,0,0.60}
\definecolor{fzjblue2}{cmyk}{0.45,0.20,0,0}
\definecolor{fzjgray}{cmyk}{0,0,0,10}
\definecolor{fzjyellow}{cmyk}{0.05,0,0.75,0}
\definecolor{fzjred}{cmyk}{0,0.75,0.40,0}
\definecolor{fzjgreen}{cmyk}{0.35,0,0.75,0}
\definecolor{fzjviolet}{cmyk}{0.35,0.55,0,0}
\definecolor{fzjorange}{cmyk}{0,0.35,0.70,0}

\newcommand{\Appref}[1]{Appendix~\ref{sec:#1}}

\newcommand{\Figref}[1]{Figure~\ref{fig:#1}\xspace}

\newcommand{\goesto}{\ensuremath{\rightarrow}}
\newcommand{\infinity}{\infty}

\newcommand{\one}{\ensuremath{\mathbbm{1}}}

\newcommand{\FF}{\ensuremath{\mathbb{F}}}
\newcommand{\HH}{\ensuremath{\mathbb{H}}}
\newcommand{\UU}{\ensuremath{\mathbb{U}}}
\newcommand{\PP}{\ensuremath{\mathbb{P}}}
\newcommand{\LLambda}{\ensuremath{\mathbb{\Lambda}}}

\newcommand{\oneover}[1]{\ensuremath{\frac{1}{#1}}}                             %   1/[argument]
\newcommand{\inverse}{\ensuremath{^{-1}}}                                       %   argument^-1
\newcommand{\transpose}{\ensuremath{{}^{\top}}}
\newcommand{\adjoint}{\ensuremath{{}^{\dagger}}}

\let\builtinLaTeX\LaTeX
\def\LaTeX{\builtinLaTeX\xspace}

%% file: affiliations.tex
\newcommand{\bonn}{
    Helmholtz-Institut f\"{u}r Strahlen- und Kernphysik,
    Rheinische Friedrich-Wilhelms-Universit\"{a}t Bonn, 53115 Bonn, Germany
}

\newcommand{\ikp}{
    Institut f\"{u}r Kernphysik,
    Forschungszentrum J\"{u}lich, 54245 J\"{u}lich, Germany
}

\newcommand{\ias}{
    Institute for Advanced Simulation,
    Forschungszentrum J\"{u}lich, 54245 J\"{u}lich, Germany
}

\newcommand{\jsc}{
    JARA \& J\"{u}lich Supercomputing Center,
    Forschungszentrum J\"{u}lich, 54245 J\"{u}lich, Germany
}

\newcommand{\liverpool}{
    Department of Mathematical Sciences,
    University of Liverpool, Liverpool, L69 7ZL, United Kingdom
}

\newcommand{\casa}{
    Center for Advanced Simulation and Analytics (CASA),
    Forschungszentrum Jülich, 52425 J\"{u}lich, Germany
}

%% file: section/introduction.tex
%!TEX root = ../master.tex
\section{Introduction}\label{sec:intro}

The \emph{numerical sign problem} is a major hindrance for the application of stochastic methods to certain physical systems, such as QCD at finite baryon density or electronically doped systems in strongly correlated condensed matter. 
The problem refers to the extreme cost of numerically approximating integrals arising with a highly oscillatory integrand, such as path integrals with complex-valued actions.
Because partition functions are exponential in the action, the numerical costs typically scale exponentially in the spacetime volume~\cite{Splittorff:2006fu}, pushing many physically interesting systems beyond the reach of numerical investigation.

Methods that reduce the sign problem allow us to use our limited resources more efficiently and thus extend the range of systems we can investigate. 
In cases when the offending term in the Hamiltonian that induces the complex phase is small, one can rely on simple reweighting.
For small systems or ground-state properties one can forego stochastic simulations and instead use direct methods, such as tensor networks~\cite{Schneider:2021}.
Complex Langevin is another popular method to fight the sign problem, but a lot of technology is required to guarantee it converges to the right distribution~\cite{kogut2019,berger2021complex}. 
Each of these methods have their own limitations, by no means fully solving the sign problem.

Here we focus on \emph{contour deformation} to alleviate the sign problem.
One transforms the integration domain of the path integral to a more favorable manifold in the high-dimensional complex space where the sign oscillations are reduced~\cite{Cristoforetti:2013wha,Cristoforetti:2014gsa,Lawrence:2018mve,Warrington:2019kzf,Wynen2019,Detmold:2020ncp,alexandru2020complex,berger2021complex,Detmold:2021ulb,Rodekamp2022}.
Such deformations are formally allowed as long as one does not cross any singularities of the integrand and one preserves the homology class of the integral.
There exist manifolds, so-called \emph{Lefschetz thimbles}, where the complex phase remains fixed.
In theories where one thimble dominates, the sign problem is solved since the constant complex phase on the thimble can be factored outside of the path integral.
Even when multiple thimbles contribute, each with a different but constant phase, the sign problem is not eliminated, but is expected to be improved.
While the locations of these thimbles are not known \emph{a priori} they can be found by integrating holomorphic flow equations. 
Unfortunately the numerical determination of the complete set of contributing thimbles is quite costly: mapping out their full constellation is just as difficult as the original sign problem.

As the goal in our work is to \emph{alleviate} the sign problem (as opposed to eliminating it), a natural question arises:  given finite computational resources, which contour deformations are most efficient for the problem at hand at alleviating the sign problem \emph{sufficiently}, meaning that observables can be extracted in a statistically meaningful and reliable manner.
In previous studies of the Hubbard model~\cite{Wynen2020,Rodekamp2022} we trained neural networks (NNs) on flowed configurations to parametrize an integration manifold called a \emph{learnifold}~\cite{Alexandru2016,alexandru2020complex}. 
This deformation worked very well at alleviating the sign problem for various doped Hubbard systems. 
In particular, we provided physical results of a doped Hubbard model for carbon nano-systems up to 18 ion sites~\cite{Rodekamp2022}.
However this method still comes at the cost of generating flowed training data and training a neural network.
Here we study the simplest imaginable deformation: shifting the integration manifold by a global imaginary constant offset.
We find that optimizing the offset substantially reduces the computational demands and often yields a contour deformation of equal potency.

A constant offset induces no Jacobian; keeping the method simple.
Further, a constant offset does not require modification of the Monte Carlo algorithm, nor does it require generation of training data and training of NNs.
In ref.~\cite{Alexandru2016}, for example, it was shown for the Thirring model that a calculation on the \emph{tangent plane}, a constant offset that intersects the classical saddle point of the main Lefschetz thimble, is sufficient at alleviating the sign problem.
We have also used this deformation as a comparison to our neural networks in previous publications \cite{Wynen2020,Rodekamp2022}. 
For the Hubbard model, however, we find that for certain values of the chemical potential, the tangent plane does not meaningfully alleviate the sign problem.
We will show how to incorporate quantum corrections to the saddle point, thereby obtaining a better constant shift that corresponds to the effective action obtained by inclusion of 1-particle irreducible terms.
Even then there are cases where we resort to numerical optimization of the imaginary offset.

We study the fermionic Hubbard model.
As opposed to the systems investigated in our earlier work~\cite{Rodekamp2022}, here we also consider systems that are non-bipartite, such as the fullerenes $C_{20}$ and $C_{60}$.  
In the following section we provide the formal aspects of our method, providing derivations for the location of the classical and quantum-corrected saddle points that we use to determine our constant offsets.
In \cref{sec:method} we continue with the description of our method for numerically determining the optimized plane.
We then demonstrate the efficacy of our methods by providing numerical results of various Hubbard systems in \cref{sec:results}.
We recapitulate in \cref{sec:conc}.  
To keep the presentation reasonable, we place formal (and tedious) derivations in the appendices.

%% file: section/theory.tex
%!TEX root = ../master.tex
\section{Formalism}\label{sec:theory}

\subsection{The Hubbard Model}

The Hubbard model describes the interacting behavior of particles on a lattice. In our case these are electrons on a lattice of ions.
It consists of a tight binding term and an onsite interaction representing electron-electron repulsion.  
It takes into account external influences on the overall particle number, like doping or an applied voltage, via a chemical potential $\mu$\cite{Hubbard1959,Hubbard1963,Hubbard1964,Hubbard1964a,Hubbard1965,Hubbard1967,Hubbard1967a}. We formulate our theory in the \emph{particle-hole basis}~\cite{Brower:2011av}
\begin{align}
	\label{eq:H and q}
	H &= -{\kappa}\sum_{\langle x,y \rangle}\left(a^\dagger_{x}a_{y}^{} - b^\dagger_{x}b_{y}^{}\right) + \frac{U}{2}\sum_{x} q_x^2 - {\mu} \sum_x q_x
	&
	q_x &= a\adjoint_x a_x - b\adjoint_x b_x
\end{align}
where $\kappa$ is the hopping parameter of neighboring lattice sites, $U$ provides the strength of interaction of two electrons sharing one lattice site, and $q$ is the local charge operator relative to half filling. Alternatively the sum over neighboring sites can be represented with the hopping matrix $K$ which is $\kappa$ times the adjacency matrix of the lattice. This option also allows for individual hopping parameters.
The sum in $x$ is over all $N_x$ sites.
The $a^{}$ $(a^{\dagger})$ operator implements particle destruction (creation) and $a^{\dagger}a$ counts particles.
Similarly the $b^{}$ $(b^{\dagger})$ operators destroy (create) holes and $b^{\dagger} b$ counts them.

\subsection{Finite lattices considered in this work}

\begin{figure}[h]
	\begin{subfigure}[c]{0.22\linewidth}
		\centering
		\resizebox{\textwidth}{!}{
			\begin{tikzpicture}
				\colorlet{edge}{black}
				\colorlet{back}{white}
				\draw[fill=black,thick] (0,0) circle (4pt);
				\draw[draw=edge, line width=0.6, line cap=round] (0, 0) -- (1, 0);
				\draw[fill=black,thick] (1,0) circle (4pt);
				\draw[draw=edge, line width=0.6, line cap=round] (1, 0) -- (1.5, 0.8660);
				\draw[fill=black,thick] (1.5, 0.8660) circle (4pt);
				\draw[draw=edge, line width=0.6, line cap=round] (1, 0) -- (1.5, -0.8660);
				\draw[fill=black,thick] (1.5, -0.8660) circle (4pt);
				\draw[draw=edge, line width=0.6, line cap=round] (1.5, -0.8660) -- (2.5, -0.8660);
				\draw[fill=black,thick] (2.5, -0.8660) circle (4pt);
				\draw[draw=edge, line width=0.6, line cap=round] (1.5, 0.8660) -- (2.5, 0.8660);
				\draw[fill=black,thick] (2.5, 0.8660) circle (4pt);
				\draw[fill=black,thick] (3.0, 0) circle (4pt);
				\draw[draw=edge, line width=0.6, line cap=round] (3, 0) -- (4, 0);
				\draw[fill=black,thick] (4.0, 0) circle (4pt);
				\draw[draw=edge, line width=0.6, line cap=round] (3, 0) -- (2.5, 0.8660);
				\draw[draw=edge, line width=0.6, line cap=round] (3, 0) -- (2.5, -0.8660);
			\end{tikzpicture}
		}
		\subcaption[8-site lattice]{8-site}
		\label{fig:8siteHex}
	\end{subfigure}
	\begin{subfigure}[c]{0.22\linewidth}
		\centering
		\resizebox{\textwidth}{!}{
			\begin{tikzpicture}
				\colorlet{edge}{black}
				\colorlet{back}{white}
				\draw[draw=edge, line width=0.6, line cap=round] (0, 0) -- (1, 0);
				\draw[draw=edge, line width=0.6, line cap=round] (1, 0) -- (1.5, 0.8660);
				\draw[draw=edge, line width=0.6, line cap=round] (1.5, 0.8660) -- (2.5, 0.8660);
				\draw[draw=edge, line width=0.6, line cap=round] (2.5, 0.8660) -- (3, 1.7321);
				\draw[draw=edge, line width=0.6, line cap=round] (3, 1.7321) -- (4, 1.7321);
				\draw[draw=edge, line width=0.6, line cap=round] (4, 1.7321) -- (4.5, 0.8660);
				\draw[draw=edge, line width=0.6, line cap=round] (4.5, 0.8660) -- (5.5, 0.8660);
				\draw[draw=edge, line width=0.6, line cap=round] (5.5, 0.8660) -- (6, 0);
				\draw[draw=edge, line width=0.6, line cap=round] (6, 0) -- (7, 0);
				\draw[draw=edge, line width=0.6, line cap=round] (1, 0) -- (1.5, -0.8660);
				\draw[draw=edge, line width=0.6, line cap=round] (1.5, -0.8660) -- (2.5, -0.8660);
				\draw[draw=edge, line width=0.6, line cap=round] (2.5, -0.8660) -- (3, -1.7321);
				\draw[draw=edge, line width=0.6, line cap=round] (3, -1.7321) -- (4, -1.7321);
				\draw[draw=edge, line width=0.6, line cap=round] (4, -1.7321) -- (4.5, -0.8660);
				\draw[draw=edge, line width=0.6, line cap=round] (4.5, -0.8660) -- (5.5, -0.8660);
				\draw[draw=edge, line width=0.6, line cap=round] (5.5, -0.8660) -- (6, 0);
				\draw[draw=edge, line width=0.6, line cap=round] (3, 0) -- (4, 0);
				\draw[draw=edge, line width=0.6, line cap=round] (3, 0) -- (2.5, 0.8660);
				\draw[draw=edge, line width=0.6, line cap=round] (3, 0) -- (2.5, -0.8660);
				\draw[draw=edge, line width=0.6, line cap=round] (4.5, 0.8660) -- (4, 0);
				\draw[draw=edge, line width=0.6, line cap=round] (4.5, -0.8660) -- (4, 0);
				\draw[fill=black,thick] (0, 0) circle (4pt);
				\draw[fill=black,thick] (1, 0) circle (4pt);
				\draw[fill=black,thick] (3, 0) circle (4pt);
				\draw[fill=black,thick] (4, 0) circle (4pt);
				\draw[fill=black,thick] (6, 0) circle (4pt);
				\draw[fill=black,thick] (7, 0) circle (4pt);
				\draw[fill=black,thick] (1.5, 0.8660) circle (4pt);
				\draw[fill=black,thick] (2.5, 0.8660) circle (4pt);
				\draw[fill=black,thick] (4.5, 0.8660) circle (4pt);
				\draw[fill=black,thick] (5.5, 0.8660) circle (4pt);
				\draw[fill=black,thick] (1.5, -0.8660) circle (4pt);
				\draw[fill=black,thick] (2.5, -0.8660) circle (4pt);
				\draw[fill=black,thick] (4.5, -0.8660) circle (4pt);
				\draw[fill=black,thick] (5.5, -0.8660) circle (4pt);
				\draw[fill=black,thick] (3, 1.7321) circle (4pt);
				\draw[fill=black,thick] (4, 1.7321) circle (4pt);
				\draw[fill=black,thick] (3, -1.7321) circle (4pt);
				\draw[fill=black,thick] (4, -1.7321) circle (4pt);
			\end{tikzpicture}
		}
		\subcaption[18-site lattice]{18-site}
		\label{fig:18siteHex}
	\end{subfigure}
	\begin{subfigure}[c]{0.22\linewidth}
	\centering
	\includegraphics[width=\textwidth]{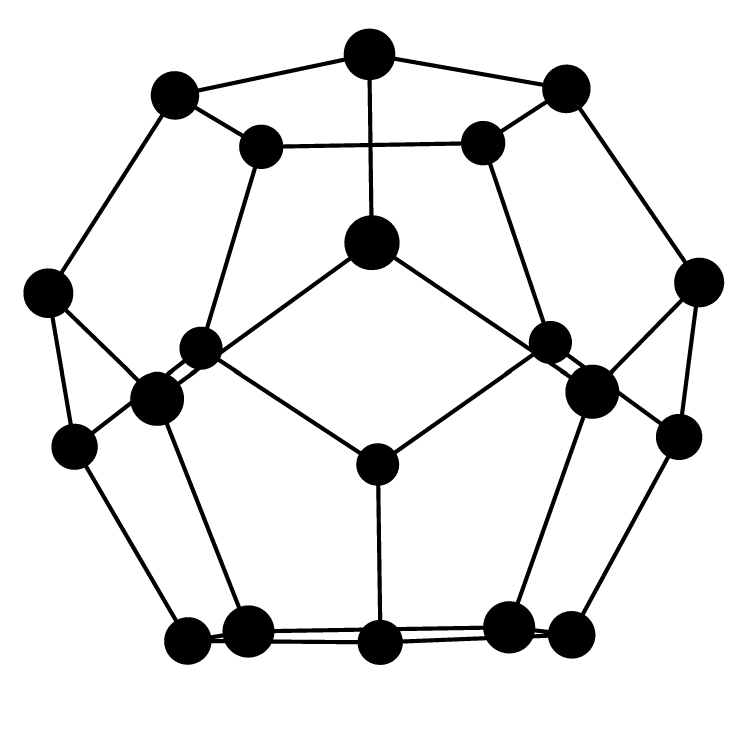}
	\subcaption{$C_{20}$}
	\label{fig:c20}
\end{subfigure}
\begin{subfigure}[c]{0.22\linewidth}
	\centering
	\includegraphics[width=\textwidth]{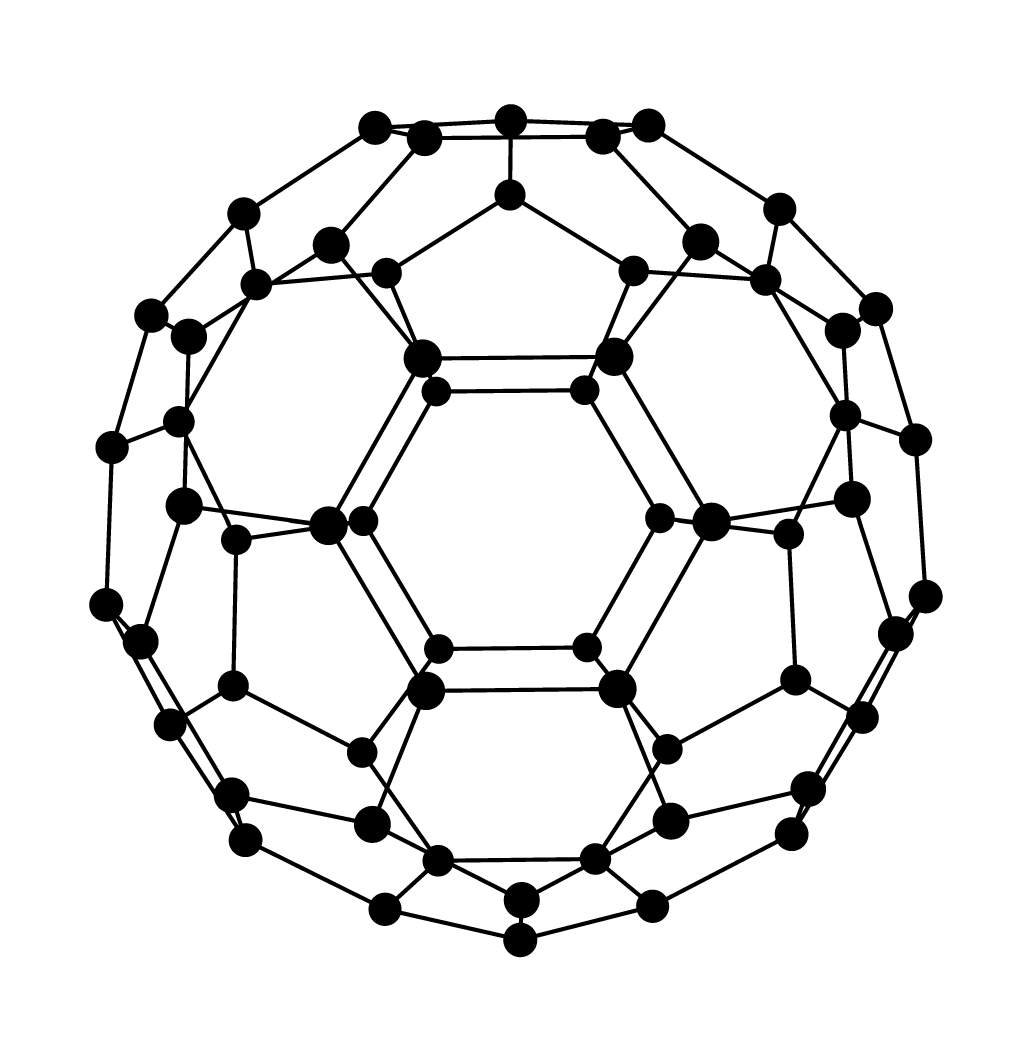}
	\subcaption{$C_{60}$}
	\label{fig:c60}
\end{subfigure}
\caption{Spatial lattices considered in this paper.}\label{fig:lattices}
\end{figure}
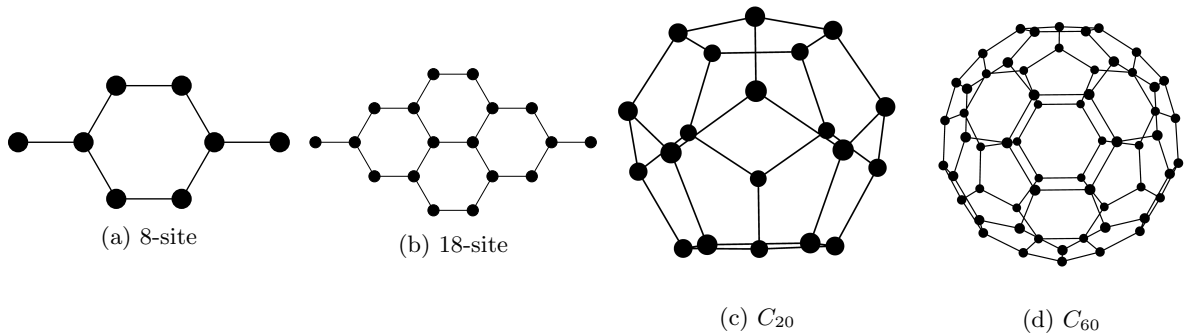

As we explain below, stochastic simulations of the Hubbard model suffer from the sign problem when the geometry of the system is non-bipartite and/or a non-zero chemical potential is present. A lattice is \emph{bipartite} when its sites can be divided into two groups, such that each site has only neighbors of the other group. Another way to think of it is that each closed path must traverse an even number of links. 
In this paper we will investigate both cases, with the 8- and 18-site honeycomb lattices as bipartite examples and the $C_{20}$ and $C_{60}$ fullerenes as non-bipartite examples. 
The 8- and 18-site honeycomb lattices consist of $2\times2$ and $3\times3$ unit cells respectively and in this work are assumed to have periodic boundary conditions. 
$C_{20}$ is a dodecahedron with 12 equal pentagons. 
$C_{60}$ is a truncated icosahedron with 12 pentagons and 20 hexagons.
The four lattice structures are visualised in \cref{fig:lattices}.
All of the lattices we consider are \emph{site transitive}, meaning that symmetries of the lattice can map any site to any other site, an analog to translation invariance.

\subsection{The path-integral formulation of the Hubbard model}

The expectation value of any quantum mechanical operator $\hat{O}$ can be calculated within the path integral formalism,
\begin{equation}\label{eq:expval}
	\expval{\hat{O}} = \frac{1}{\mathcal{Z}} \int \mathcal{D}\phi\,  \hat{O}\left[ \phi \right]e^{-S\left[\phi\right]}
\end{equation}
where the partition function is
\begin{equation}\label{eq:partf}
	\mathcal{Z} = \int \mathcal{D}\phi\, e^{-S\left[\phi\right]}\
\end{equation}
and $\mathcal{D}\phi = \lim_{N_t\rightarrow\infty}\prod_{x}^{N_x}\prod_{t}^{N_t}d\phi_{x,t}$ and $S[\phi]$ is the action that defines the system. 
It is common practice to estimate expectation values \eqref{eq:expval} with importance sampling
\begin{equation}\label{eq:MC}
	\int \mathcal{D}\phi\,  \hat{O}\left[ \phi \right]\mathbb{P}\left[\phi\right] = \lim_{N\rightarrow\infinity} \oneover{N}\sum_i^N \hat{O}\left[ \phi_i \right] , \quad \phi_i\sim \mathbb{P}\left[\phi_i\right],
\end{equation}
drawing configurations $\phi$ from the probability distribution $\mathbb{P}$ by Markov Chain Monte Carlo (MCMC) methods such as Hybrid (or Hamiltonian) Monte Carlo (HMC)~\cite{Duane1987}.
For purely real actions it is straight forward to choose $\mathbb{P}\left[\phi\right] = e^{-S\left[\phi\right]}/\mathcal{Z}$.   
For generally complex actions $S[\phi]=S_R[\phi]+i S_I[\phi]$, however, one typically separates the complex phase and absorbs it into the definition of the observable,
\begin{equation}\label{eq:reweighting}
	\expval{\hat{O}} = \frac{\expval{\hat{O}\exp(-i S_I)}_R}{\expval{\exp(-i S_I)}_R}
\end{equation}
where the subscript \emph{R} indicates sampling according to the probability defined by the real part of the action, $\mathbb{P}[\phi]=e^{-S_R[\phi]}/\mathcal{Z}_R$.
This process is called \emph{reweighting} and it exactly produces correct expectation values \eqref{eq:expval} in the limit of infinite statistics.

However, for finite statistics, the oscillating phase in the denominator of the reweighting \eqref{eq:reweighting}, known as the \emph{average phase},
\begin{equation}\label{eq:average_phase}
	\expval{\exp(-i S_I)}_R = \dfrac{\int \mathcal{D}\phi\,  e^{-iS_I\left[\phi\right]}e^{-S_R\left[\phi\right]}}{\int \mathcal{D}\phi\,  e^{-S_R\left[\phi\right]}}=\frac{\mathcal{Z}}{\mathcal{Z}_R}
\end{equation} 
can be hard to numerically estimate.
Stronger oscillation and its attendant cancellations become more severe for larger system sizes and some parameters. 
We call the average phase's absolute value
\begin{equation}\label{eq:sp}
	\Sigma=\abs{\expval{\exp(-i S_I)}_R} = \abs{\frac{\mathcal{Z}}{\mathcal{Z}_R}}
\end{equation} 
the \emph{statistical power} and we use it to quantify the sign problem. 
When the statistical power is 1, the path integral is sign-problem free; a value of 0 indicates the worst possible sign problem.
While in a problem-free case the stochastic uncertainties of expectation values from an ensemble of $N_\text{cfg}$ generated configurations scale with $N^{-1/2}_\text{cfg}$, with a sign problem the statistical power effectively reduces the contribution of each configuration: the error scales with the square root of the effective number of samples~\cite{berger2021complex}, 
\begin{equation}\label{eq:neff}
	N_{\text{eff}} = \Sigma^2 \times N_{\text{cfg}} \ .
\end{equation}

Trotterizing the thermal partition function $\tr{e^{-\beta H}}$, linearizing the interaction with a Hubbard-Stratonovich transformation with an integral over auxiliary fields $\phi$, and inserting resolutions of the identity in terms of Grassmann coherent states, yields an action
\begin{equation}\label{eq:action}
	S\left[\phi, \tilde{K}, \tilde{\mu} \right]= \frac{1}{2\tilde{U}} \sum_{t,x}\phi^2_{x,t} - \log\det \left( M\left[+\phi, +\tilde{K}, +\tilde{\mu}\right] \right) - \log\det \left(M\left[-\phi, -\tilde{K}, -\tilde{\mu} \right] \right) 
\end{equation}
where the dimensionless parameters $\tilde U = U\times\delta$, $\tilde \kappa=\kappa\delta$ etc.\@ are rescaled by the temporal lattice spacing $\delta=\beta/N_t$.  
The fermion matrices $M$ encode the particle and hole fermion loops exactly.
There are a variety discretizations of the fermion matrix that become exact and equal in the continuum limit~\cite{Wynen2019,Beyl:2017kwp,Brower:2011av}.
We use the `exponential' discretization in the language of ref.~\cite{Wynen2019} 
\begin{equation}\label{eq:expdis}
	M_{x't',x t}[\pm\phi,\pm K,\pm\mu]\equiv M_{x't',x t}[\pm; \phi]=\delta_{x',x}\delta_{t',t}-[\exp(\pm\tilde K)]_{x',x}e^{\pm (-i\phi_{x,t}+\tilde \mu)}B_{t'}\delta_{t',t+1} 
\end{equation}
where $B_{N_t-1}$ carries an extra $-1$ encoding the fermionic temporal antiperiodic boundary conditions.
We do not consider other discretizations in this work.
We note, however, that when $\mu\ne0$, this discretization does not suffer from ergodicity issues described in refs.~\cite{Beyl:2017kwp,Wynen2019}.

We now consider deforming our original integral by complexifying the auxiliary field $\phi$ and deforming the integration manifold.
Cauchy's theorem guarantees that this deformation leaves all holomorphic observables the same, as long as the deformation does not cross any singularities and the deformation preserves the homology class.
The statistical power depends on the imaginary part of the action weighted by its real part and thus is not holomorphic, so it is manifold-dependent.
Some manifolds may tame the oscillations, especially when they resemble Lefschetz thimbles, high-dimensional analogs of contours of steepest descent~\cite{Ulybyshev:2019fte,alexandru2020complex}. 
We have previously trained neural networks to learn the results of the holomorphic flow in a computationally tractable way~\cite{Wynen2020,Rodekamp2022}. 
An even simpler deformation, that of a constant imaginary shift in all components of $\phi$, can lead to significant alleviation of the sign problem while incurring no additional costs to the HMC algorithm. 
In particular, ref.~\cite{alexandru2020complex} showed that a constant shift that intersected the saddle point of the main thimble, producing the so called `tangent plane', sufficiently reduced the sign problem in simulations of the Thirring model.
We now consider the same constant shift to the tangent plane of the Hubbard model.

\subsection{The tangent plane of the Hubbard model}

The holomorphic flow of a configuration  $\phi$ is its image under evolution in a fictitous time $t$ by
\begin{align}
	\frac{d\phi}{dt} = ( \partial_\phi S )^*.
	\label{eq:flow}
\end{align}
The saddle point that fixes the tangent plane is found by flowing the $\phi=0$ translationally-invariant configuration to its fixed point.
Because it is a fixed point the time derivative vanishes and the saddle point $\phi_c$ satisfies
\begin{align}
	\left.\partial_\phi S[\phi]\right.\vline_{\phi=\phi_c}=0\ .
	\label{eq:fixed-point}
\end{align}
In the graphene case this saddle point has the greatest weight~\cite{Ulybyshev:2022kxq} on the semimetal side of the quantum critical point at $U\lesssim \num{3.8}$~\cite{Assaad:2013xua,ParisenToldin:2014nkk,Otsuka:2015iba,Buividovich:2018crq,Buividovich:2018yar,Ostmeyer:2020uov,Ostmeyer:2021efs}.
The saddle point $\phi_c$ has zero real part and, because the lattices we consider are site-transitive, constant imaginary part which is non-zero when $\mu\ne 0$ on bipartite lattices and generically on non-bipartite lattices. 

Leveraging the simplicity of $\phi_c=i\phi_0$ independent of space and time we can calculate the action
\begin{align}
	S[\phi_c = i \phi_0] = \oneover{2\tilde{U}} N_x N_t \phi_0^2 - \log\det\left( \one + e^{N_t \phi_0 + \beta \mu} e^{+\beta K}\right) - \log\det \left(\one + e^{- N_t \phi_0 - \beta \mu} e^{-\beta K}\right)
\end{align}
where we used the Schur complement to simplify the fermion determinants and used the spatial independence of $\phi_0$ to commute the auxiliary field terms past the hopping terms.
Using $\log \det = \tr \log$, transforming to the basis where the hopping matrix is diagonalized, and requiring $\phi_c$ to be a fixed point \eqref{eq:fixed-point} leads to
\begin{align}
	\label{eq:transcendental}
	\phi_0/\delta = -\frac{U}{N_x}\sum_{k}\tanh\left(\frac{\beta}{2}\left[\epsilon_k+\mu+\phi_0 /\delta\right]\right)
\end{align}
where the sum is over the $N_x$ modes of the hopping matrix $K$ with noninteracting energy eigenvalue $\epsilon_k$; we provide a detailed derivation in \Appref{appendix tangent plane}.
Writing $\delta=\beta/N_t$ shows that this transcendental equation contains only temporal continuum quantities, except for the combination $\phi_0 N_t$, which will stay fixed as we go towards the time continuum limit $N_t\goesto\infty$.
We see that $\phi_0/\tilde{U}$ is bounded between $-1$ and $+1$ and can cheaply determine the imaginary offset of the tangent plane $\phi_0$ solving this equation numerically.
Figure~\ref{fig:trans18} shows the behavior of the tangent plane for the 18-site honeycomb problem as a function of $\mu$ for select values of $U$ and $\beta$. For large $\beta$ where the $\tanh$ becomes a sign function, the tangent plane has plateaus that are connected by constant slopes. The location of those depends on the spectrum of the hopping matrix $K$; an example is shown in fig.~\ref{fig:trans18}.

\subsubsection{Properties of the tangent plane in the $\mu=0$, $\beta\to\infty$ limit}

The imaginary offset vanishes $\phi_0=0$ so that the tangent and real planes coincide if the noninteracting energy eigenvalues $\epsilon_k$ are symmetric about zero.
This symmetry naturally occurs for bipartite lattices, such as the honeycomb lattice, in the absence of chemical potential $\mu=0$.
For these cases $\phi_0=0$ for any inverse temperature $\beta$.

\begin{figure}[h]
	\centering
	\includegraphics[width=.8\textwidth]{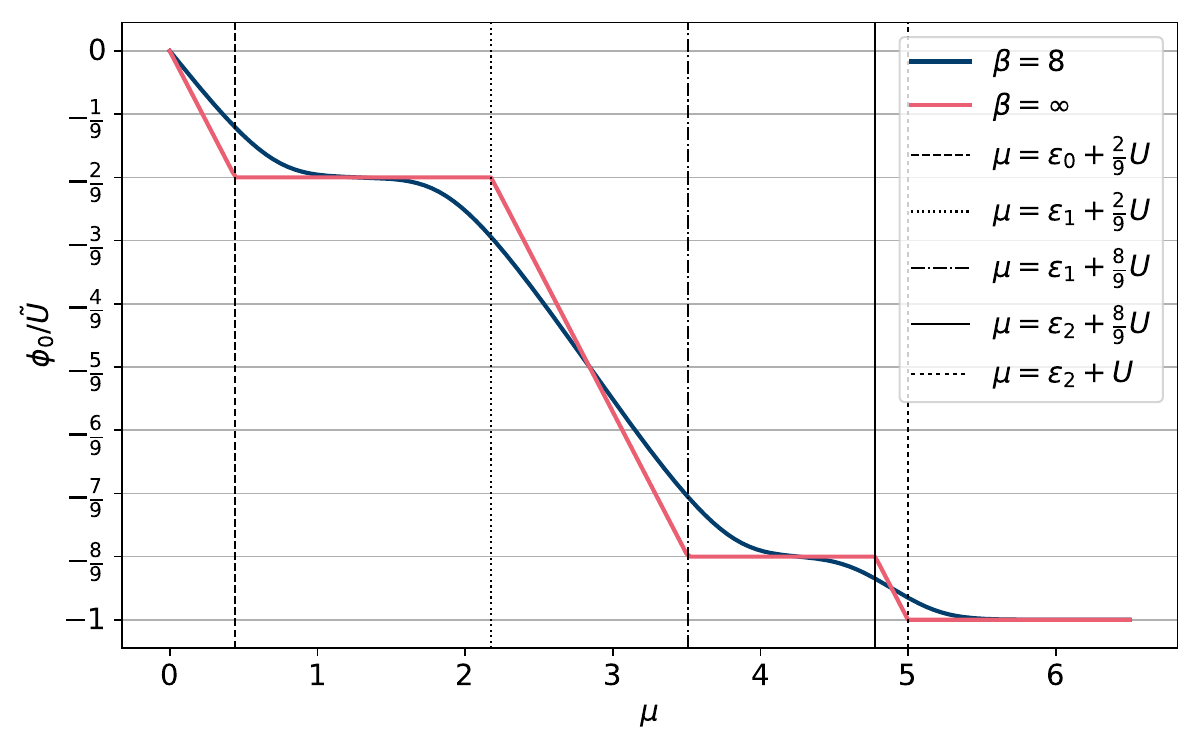}
	\caption{Tangent plane of 18-site honeycomb lattice. The vertical lines mark transitions where the argument of a $\tanh$ that determines the tangent plane \eqref{eq:transcendental} switches its sign. At the beginning of a downwards slope in $\beta\to\infty$ the sign switches from $-1$ to $0$ and at the end from $0$ to $+1$.
	}
	\label{fig:trans18}
\end{figure}

Moreover, in the $\beta\to\infty$ limit the $\tanh$ functions become the sign function, and if there are equal numbers of positive and negative noninteracting eigenvalues (not necessarily symmetric about zero), the sum also vanishes and the tangent plane again corresponds to the real plane.
Both $C_{20}$ and $C_{60}$ have non-symmetric spectra but $C_{60}$ enjoys equally many positive and negative noninteracting energies.
In \Figref{trans18} we show how finite temperature smooths the piecewise-linear $\beta=\infty$ tangent plane for the 18-site honeycomb lattice.

\begin{figure}[h]
	\begin{subfigure}[t]{0.45\linewidth}
		\includegraphics[width=\linewidth]{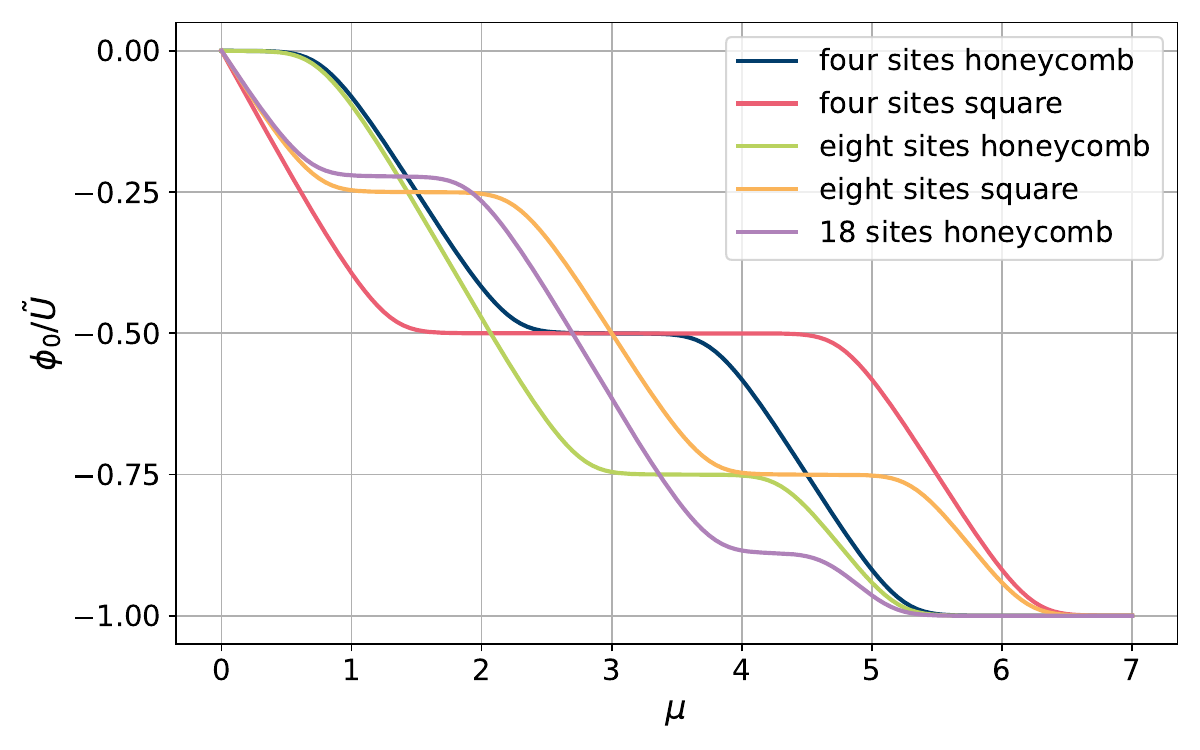}
		\subcaption{Comparison of different bipartite lattices at $U=2$ and $\beta=10$. For the bipartite lattices the tangent plane is antisymmetric around $\mu=0$.}
	\end{subfigure}
	\begin{subfigure}[t]{0.45\linewidth}
		\includegraphics[width=\linewidth]{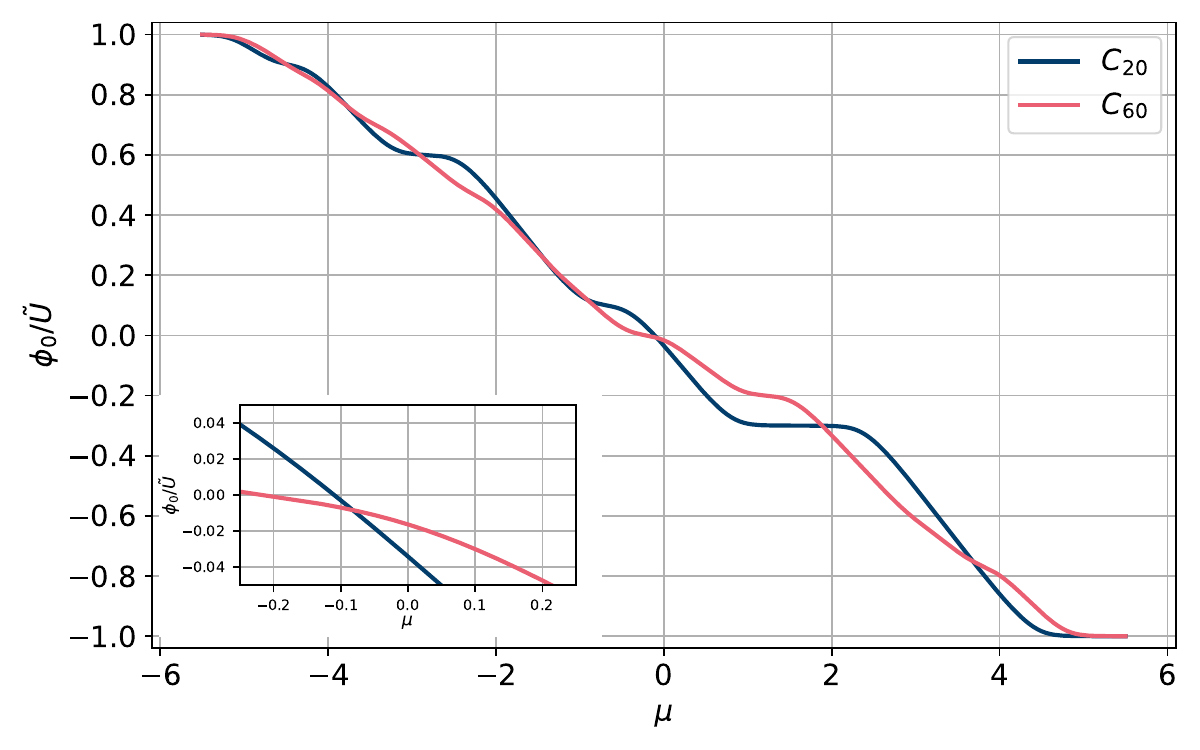}
		\subcaption{Comparison of the two fullerenes we investigated at $U=2$ and $\beta=10$. For the non-bipartite lattices the tangent plane is not symmetric.}
	\end{subfigure}\\
	\begin{subfigure}[t]{0.45\linewidth}
		\includegraphics[width=\linewidth]{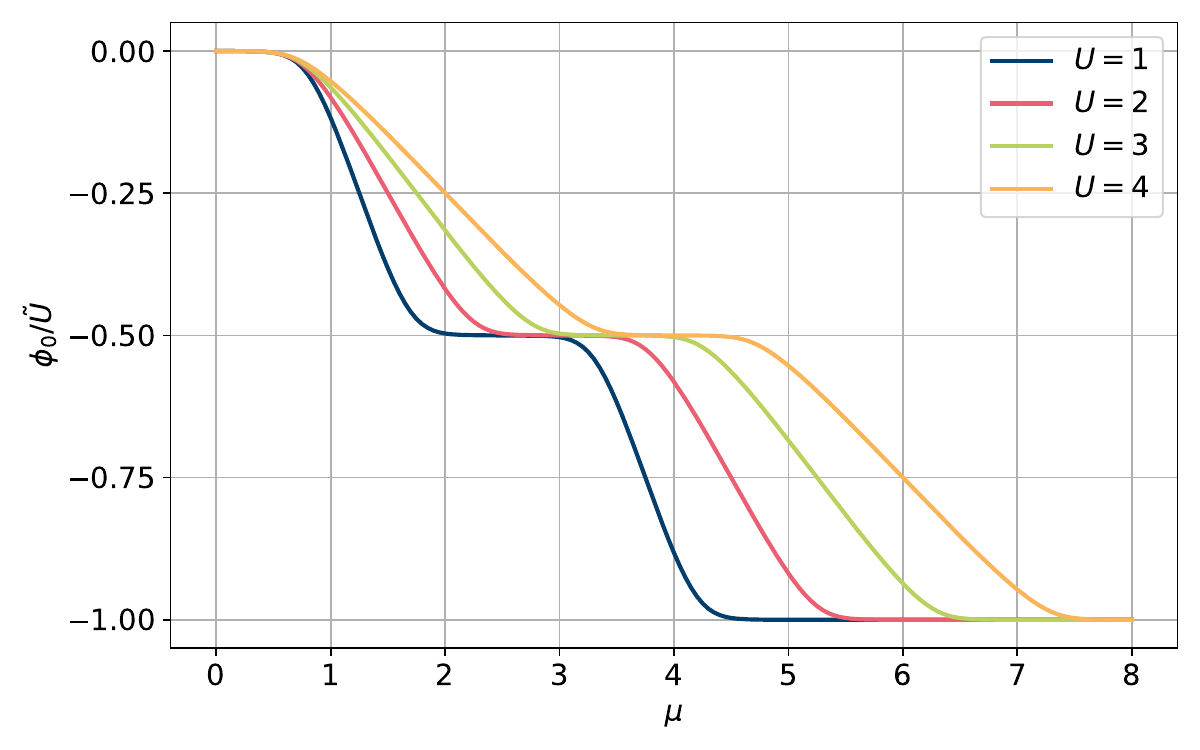}
		\subcaption{The four site honeycomb lattice at $\beta=10$ for different $U$.}
	\end{subfigure}
	\begin{subfigure}[t]{0.45\linewidth}
		\includegraphics[width=\linewidth]{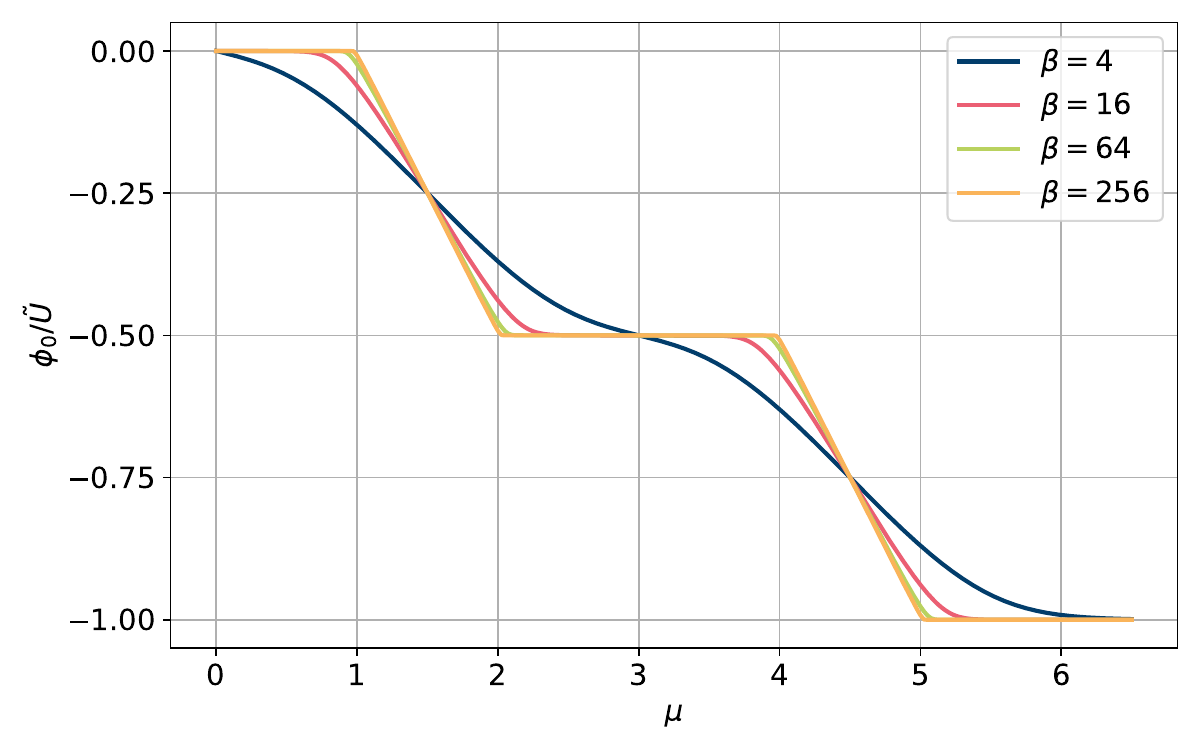}
		\subcaption{The four site honeycomb lattice at $U=2$ for different $\beta$.}
	\end{subfigure}
	\caption{Tangent plane offsets (normalized to $\tilde{U}$) depending on the chemical potential $\mu$, for various systems and parameters.}
	\label{fig:tangentplanes}
\end{figure}

In \cref{fig:tangentplanes} we compare the behavior of the tangent plane for varying $U$, $\mu$, $\beta$, and different lattices, both bipartite and non-bipartite.
The results at $\mu=0$ shown in this figure confirm our statements above.  
For the fullerene results, which are non-bipartite, the choice of $\beta=10$ is large enough that the resulting tangent plane at $\mu=0$ is nearly identical to the real plane. 

\subsubsection{Properties of the tangent plane in the $\mu\beta\to\infty$ limit}
Another interesting scenario is to consider the behavior of the tangent plane in the $\mu\beta\to\infty$ limit.  
To understand the behavior in this limit we start again with the action \eqref{eq:action}.
With repeated application of Schur's complement, we can express (see ref.~\cite{Wynen2019} for an explicit derivation)
\begin{align}\label{eq:MU}
	\log\det M[\pm]&=\log\det\left(\one+\mathbb{F}[\pm]\right)
	&
	\mathbb{F}[\pm]=\prod_{t=0}^{N_t-1}[e^{\pm\tilde K}][e^{\mp i\phi_t}]e^{\pm\tilde\mu}\ .
\end{align}
where the 0th timeslice is rightmost and each term in $[$square brackets$]$ in the product represents a space$\times$space matrix. Since the chemical potential term is proportional to the identity, we can bring it out of the product, so the action becomes
\begin{equation}\label{eq:alternative_action}
	S[\bm\phi]=\frac{\bm\phi^2}{2\tilde U}-\log\det\left(\one+e^{-\mu\beta}\prod_{t=0}^{N_t-1}[e^{-\tilde K}][e^{ i\phi_t}]\right)-\log\det\left(\one+e^{+\mu\beta}\prod_{t=0}^{N_t-1}[e^{\tilde K}][e^{- i\phi_t}]\right)\ .
\end{equation}
In the limit of asymptotically large $\beta\mu$ the determinants simplify
\begin{align}
\lim_{\mu\beta\to\infty}S[\phi]
	=&\frac{\bm\phi^2}{2\tilde U}-\log\det\left(\one\right)-\log\det\left(e^{+\mu\beta}\prod_{t=0}^{N_t-1}[e^{\tilde K}][e^{- i\phi_t}]\right)+\mathcal{O}(e^{-\mu\beta})\nonumber\\
	=&\frac{\bm\phi^2}{2\tilde U}-N_x\mu\beta-\log\det\left(\prod_{t=0}^{N_t-1}[e^{+\tilde K}][e^{- i\phi_t}]\right)\nonumber\\
	=&\frac{\bm\phi^2}{2\tilde U}-N_x\mu\beta-\log\left[ e^{-i\Phi} \det e^{+K\beta} \right] \nonumber\\
	=&\frac{\bm\phi^2}{2\tilde U}-N_x\mu\beta + i\Phi - \beta\tr{K}
	\label{eqn:ub large limit}
\end{align}
where we define $\Phi = \sum_{x,t}\phi_{x,t}$.
Since our hopping matrices have no self hopping $\tr{K}=0$.
Now solving for critical points $\partial S/\partial\phi = 0$, we find the main critical point at constant field with $\phi_c=-i\tilde U$. This demonstrates that the tangent plane approaches $-\tilde U$ in the large $\mu\beta$ limit. 
Similarly, in the $\mu\beta\to-\infty$ limit one finds
\begin{equation}
	S[\bm\phi]=\frac{\bm \phi^2}{2\tilde U}+N_x\mu\beta-i\Phi\ ,
\end{equation}
and the tangent plane approaches $+\tilde{U}$. 

The resulting tangent plane shift in this limit has important implications for our stochastic calculations.
Simulating on the tangent plane means using components of the field, $\phi_j$, that are offset by $-i\tilde U$, i.e.\ $\phi_j\to\phi_j-i\tilde U \ \forall\ j$.
The resulting action under this deformation becomes
\begin{equation}
	S[\bm \phi-i\tilde U]=\frac{\bm \phi^2}{2\tilde U}-\frac{N_x U\beta}{2}+N_x U\beta-N_x\mu\beta
	=\left(\frac{U}{2}-\mu\right)\beta N_x+\frac{\bm\phi^2}{2\tilde U}\ .\label{eq:plug_in_shift_large_mu}
\end{equation}
This means that the action on this flat contour is purely real in this limit.  That is, the tangent plane \emph{\textbf{completely solves the sign problem}} in this limit.
It is equivalent, up to some overall shift in the energy, to a \emph{quenched} calculation, with no fermion matrix.
Later we show numerical results that confirm these findings.

\subsection{Quantum corrections to the saddle point}

The saddle point that defines the tangent plane corresponds to the critical point of the classical action.  
In quantum field theory the location of this point shifts due to the presence of quantum fluctuations which, in our case, corresponds to thermal fluctuations.
We can estimate this shift by calculating the quantum effective action and determining the extremum of this action, as is done in standard textbooks on QFT.  
This correction to the saddle point corresponds to the inclusion of all one-particle irreducible (1PI) diagrams.
Thus it represents a quantum (thermal) correction to the classical saddle point, and the ensuing constant manifold that intercepts this point is expected to reduce the sign problem. 
We assume that the maximum we find, when including higher order correction terms, will return an offset that reduces the sign-problem even more than the basic tangent plane.
To start, we assume that $\phi_{c}$ is the saddle point in the presence of quantum fluctuations and apply the saddle point approximation about this point.
That is, we expand the action in powers of a small perturbation $\eta$ about this point, $\phi=\phi_{c}+\eta$.
This gives
\begin{align}\label{eq:S_expansion}
	S[\phi_{c}+\bm\eta]=&S[\phi_{c}]+\left(\bm\eta\cdot\bm\nabla\right)S[\phi_{c}]+\frac{1}{2}\left(\bm\eta\cdot\bm\nabla\right)^2S[\phi_{c}]+\order{\eta^3} \nonumber\\
	=&S[\phi_{c}]+\left(\bm\eta\cdot\bm\nabla\right)S[\phi_{c}]+\frac{1}{2}\bm\eta\cdot\mathbb{H}_{S[\phi_{c}]}\cdot\bm\eta+\order{\eta^3} .
\end{align}
Here we have made use of the Hessian,
\begin{equation}\label{eq:definitionHessian}
	\left(\mathbb{H}_{S[\phi_{c}]}\right)_{x't',xt}=\left(\partial_{x't'}\partial_{xt^{}}S[\phi]\right)\vline_{\phi=\phi_{c}} \ .
\end{equation}
Since $\eta$ is assumed small, we will omit the $\order{\eta^3}$ terms.
Furthermore because the critical point satisfies $\eval{\nabla S[\phi]}_{\phi_{c}} = 0$ the linear terms also vanish. Hence the path integral simplifies to a Gaussian integral which we can do,
\begin{equation}
	\int \mathcal{D}\phi\,e^{-S[\phi]}\approx e^{-S[\phi_{c}]}\int \mathcal{D}\bm\eta \,e^{-\frac{1}{2}\bm\eta\cdot\mathbb{H}_{S[\phi_{c}]}\cdot\bm\eta}
	= e^{-S[\phi_{c}]}\left(\det\mathbb H_{S[\phi_{c}]}\right)^{-1/2}\equiv e^{-S_{\rm eff}[\phi_{c}]}\ .
\end{equation}
This allows us to formulate an effective action
\begin{equation}\label{eq:second_order_action}
	S_{\rm eff}[\phi_{c}]=S[\phi_{c}]+\frac{1}{2}\log\det\mathbb H_{S[\phi_{c}]}\ .
\end{equation}
The extremum of this action defines our 1PI-corrected spacetime-constant saddle point $\phi_{{c}}=i\phi_1$.
Note that without the Hessian term we recover our original action and the extremum in this case is the saddle point of our leading order classical action that defines the tangent plane \eqref{eq:transcendental}.
In comparison, our 1PI-corrected effective action \eqref{eq:second_order_action} includes the quantum effects at next to leading order (NLO).
In \Appref{appendix NLO} we show how to evaluate the Hessian \eqref{eq:definitionHessian} when $\phi_{{c}}=i\phi_1$ a spacetime constant.

\subsection{Excursion to infinite lattices}

In this section we demonstrate on select lattices how to determine the tangent plane in the infinite-volume limit.
We provide two well known examples: the 2-dimensional square and honeycomb lattices.
In the infinite volume limit we can access every mode in the first Brillouin zone (B.Z.) and we can replace the sum over noninteracting energies in \cref{eq:transcendental} with a momentum integral.
For a 2-dimensional square lattice one has
\begin{displaymath}
	\frac{1}{N_x}\sum_k\to\int_{\bm k\in B.Z.}\frac{d\bm k}{(2\pi)^2} ,
\end{displaymath}
where $\bm k\equiv (k_x,k_y)$ with $-\pi\le k_i< \pi$ (square B.Z.).
The non-interacting energies are given by 
\begin{equation}
	\epsilon(\bm k)=2\left(\cos(k_x)+\cos(k_y)\right)\ .
\end{equation}
Making these substitutions to determine the tangent plane \eqref{eq:transcendental} leads to
\begin{equation}\label{eqn:inf square}
	\phi_0/\delta=-U\int_{\bm k\in B.Z.}\frac{d\bm k}{(2\pi)^2}\tanh\left(\frac{1}{2}\beta\left[\epsilon(\bm k)+\mu+ \phi_0/\delta\right]\right)\ .
\end{equation}

For the infinite honeycomb lattice the non-orthogonal lattice translation vectors and the two-band structure means we must substitute
\begin{displaymath}
	\frac{1}{N_x}\sum_k\to\frac{3\sqrt{3}}{2}\int_{\bm k\in B.Z.}\frac{d\bm k}{(2\pi)^2}\frac{1}{2}\sum_{\sigma=\pm1}\ ,
\end{displaymath}
where the factor $3\sqrt{3}/2$ comes from the hexagonal geometry of the B.Z.\ and $\sigma$ runs over the two bands.
The non-interacting energies are~\cite{doi:10.1142/p080,CastroNeto:2007fxn}
\begin{align}
	\epsilon_{\pm}(\bm k)&=\pm|f(\bm k)|
	&
	f(\bm k)&=1+2 e^{-\frac{3 i k_x}{2}} \cos \left(\frac{\sqrt{3} k_y}{2}\right)\ .
\end{align}
So we find
\begin{equation}\label{eqn:inf graphene}
	\phi_0/\delta=-U\frac{3\sqrt{3}}{2}\int_{\bm k\in B.Z.}\frac{d\bm k}{(2\pi)^2}\frac{1}{2}\sum_{\sigma=\pm1}\tanh\left(\frac{1}{2}\beta\left[\epsilon_\sigma(\bm k)+\mu+ \phi_0/\delta\right]\right)\ .
\end{equation}
\begin{figure}[h]
\includegraphics[width=.48\textwidth]{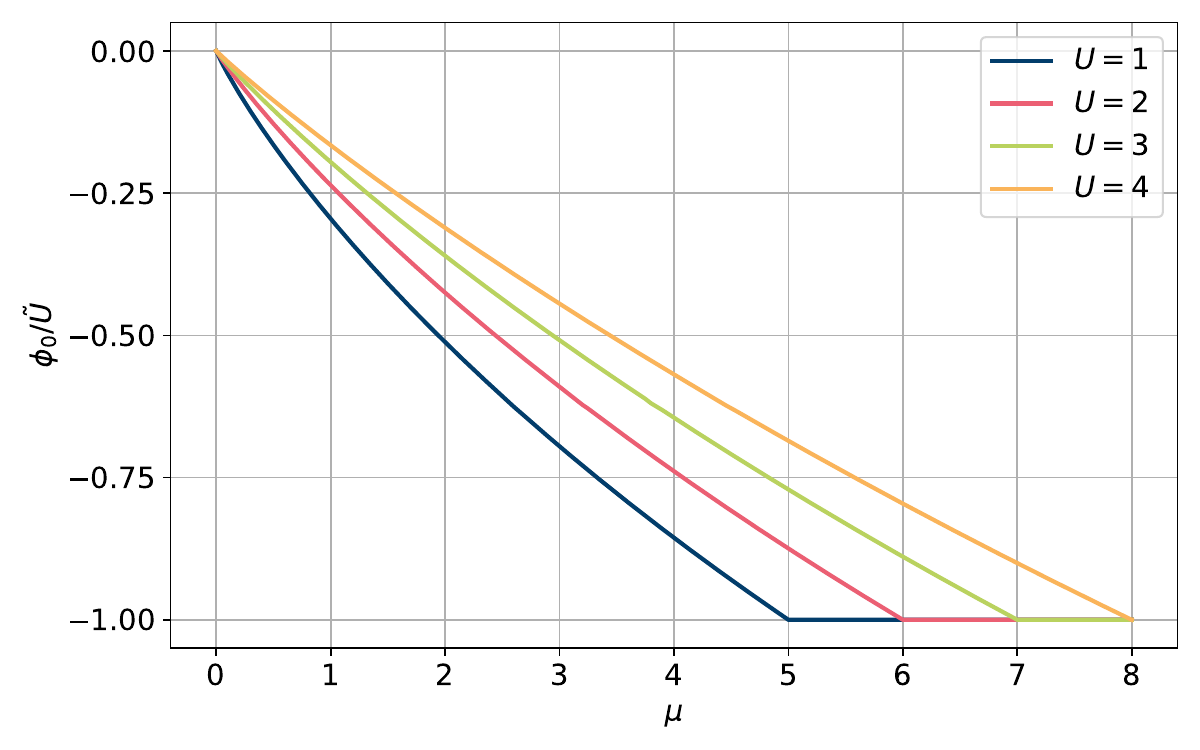}
\includegraphics[width=.48\textwidth]{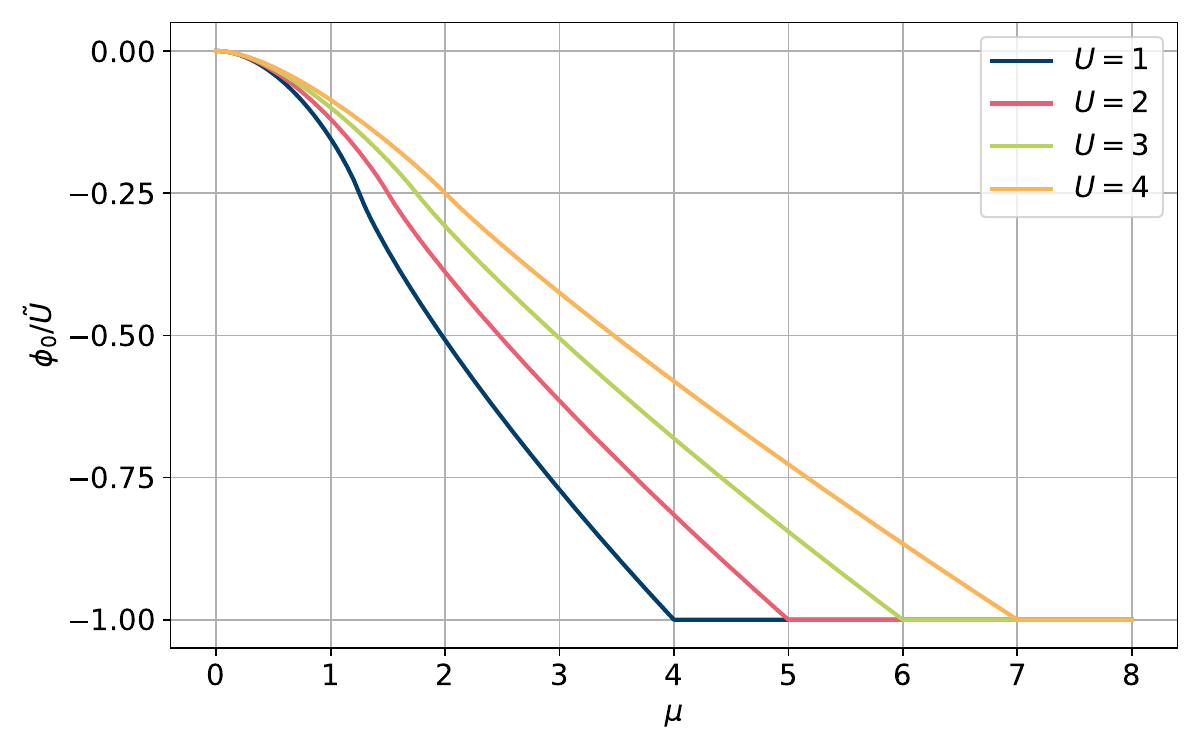}
	\caption{The zero-temperature tangent plane $\phi_0$ (normalized to $\tilde U$) for the infinite 2-D square lattice (left) and honeycomb lattice (right) as a function of chemical potential $\mu$ for various onsite interactions $U$, as labelled in the figure.}
	\label{fig:infgraphbi}
\end{figure}
We solve the square-lattice \eqref{eqn:inf square} and honeycomb \eqref{eqn:inf graphene} relations numerically.
In \cref{fig:infgraphbi} we show the solutions of $\phi_0$ for both the square and honeycomb system for select values of $U$. 
We see that $\phi_0$ remains smooth as a function of chemical potential $\mu$, even in the limit of zero temperature ($\beta\gg 1$).  
Also, in both cases, in the limit of asymptotically large $\mu$ we have $\phi_0\to -\tilde U$.

%% file: section/method.tex
%!TEX root = ../master.tex
\section{Numerical Optimization Method}\label{sec:method}

\begin{figure}[h]
	\centering
	\includegraphics[width=.8\linewidth]{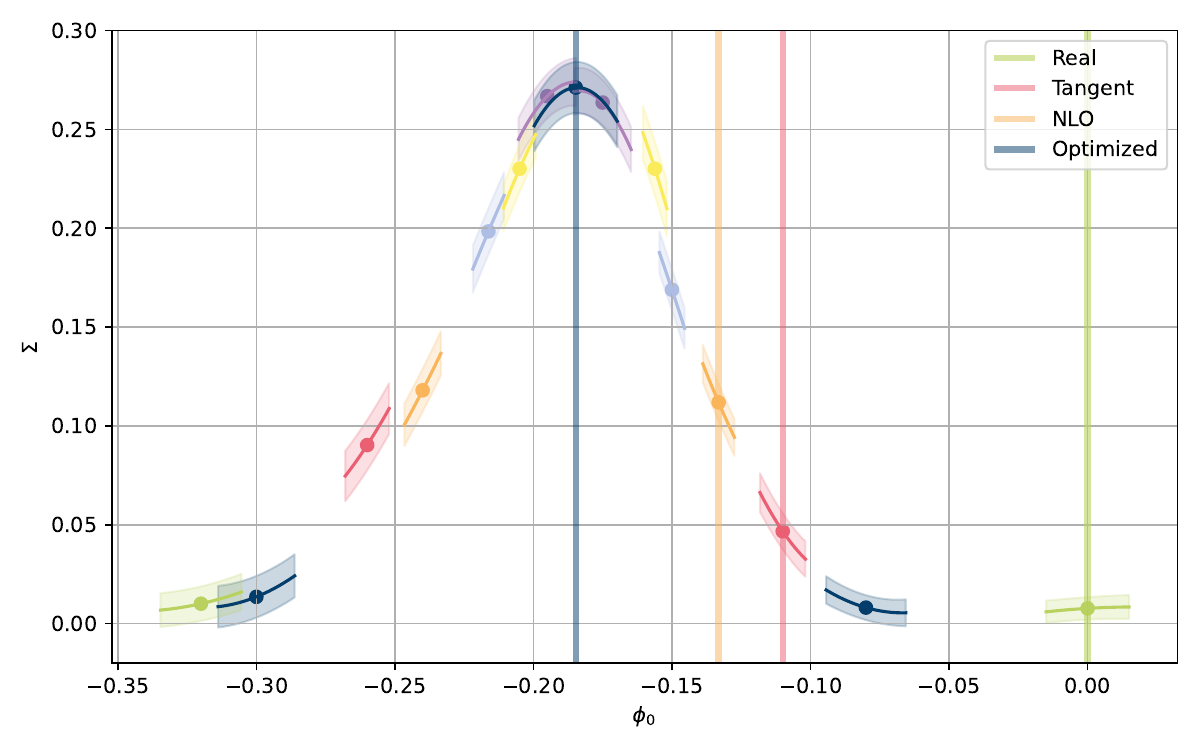}
	\caption{Visualization of effect of an imaginary offset on the sign problem. This example shows the eight site honeycomb lattice with $N_t=16$, $\beta=8$, $U=2$ and $\mu=1$. The errorbands show the bootstrap errors of the first and second derivative at the datapoint.}
	\label{fig:osp}
\end{figure}

In many cases both tangent plane and NLO offsets lead to an improvement in statistical power, with NLO typically providing modest improvement over the tangent plane (but not in all cases).  
However, a simple numerical investigation shows that one can further improve the statistical power, in most cases by shifting beyond the NLO result.  
For example, in \cref{fig:osp} we show the statistical power for the eight-site honeycomb system coming from a scan of various offsets that include the real plane, tangent plane, and the NLO offset.  
The scan shows a singular peak in the statistical power.  
However, this peak does not occur at either the tangent plane or the NLO offset.
In all our investigations of different systems to date we find similar behavior; namely, there is a singular peak in statistical power due to constant offset.
We refer to this offset that maximizes the statistical power as the \emph{optimized shift}. 

Because the greater statistical power means smaller required samples~\eqref{eq:neff}, the potential savings in computational resources when simulating at the optimized shift in comparison to either the NLO or tangent plane can be orders of magnitude. 
However, determining the location of the peak from a simple raster scan in offsets is timely and inefficient, since each point requires an HMC simulation with sufficient statistics to resolve the statistical power. 
Instead, we formulate a search algorithm like Newton-Raphson, relying on the calculation of derivatives of the statistical power using the \emph{current} HMC ensemble to make a prediction for the location of the offset that corresponds to the peak of the statistical power.
We then iterate this procedure to converge to the peak.  

Our algorithm requires the first two derivatives of the statistical power with respect to the imaginary offset $\phi_0$ from the existing Markov Chain configurations,
\begin{align}\label{eq:dSP}
	\derivative{}{\phi_0}\expval{e^{-iS_{I,\phi_0}}}_{R,\phi_0} =& \derivative{}{\phi_0}\frac{\int \mathcal{D}\phi e^{-S_{\phi_0}}}{\int \mathcal{D}\phi e^{-S_{R,{\phi_0}}}}\nonumber\\
	=&\expval{e^{-iS_{I,{\phi_0}}}}_{R,{\phi_0}}\expval{\derivative{S_{R,{\phi_0}}}{{\phi_0}}}_{R,{\phi_0}} - \expval{e^{-S_{I,{\phi_0}}}\derivative{S_{{\phi_0}}}{{\phi_0}}}_{R,{\phi_0}}
	\\
\label{eq:ddSP}
	\derivative{^2}{{\phi_0}^2}\expval{e^{-iS_{I,{\phi_0}}}}_{R,{\phi_0}} =& \derivative{}{\phi_0}\left( \expval{e^{-iS_{I,{\phi_0}}}}_{R,{\phi_0}}\expval{\derivative{S_{R,{\phi_0}}}{{\phi_0}}}_{R,{\phi_0}} - \expval{e^{-S_{I,{\phi_0}}}\derivative{S_{{\phi_0}}}{{\phi_0}}}_{R,{\phi_0}}\right) \nonumber\\
	=& \expval{e^{-iS_{I,{\phi_0}}}}_{R,{\phi_0}}\left( 2\expval{\derivative{S_{R,{\phi_0}}}{{\phi_0}}}_{R,{\phi_0}}^2+\expval{\derivative[2]{S_{R,{\phi_0}}}{{\phi_0}}-\derivative{S_{R,{\phi_0}}}{{\phi_0}}^2}_{R,{\phi_0}}\right) \nonumber\\
	&- 2\expval{e^{-S_{I,{\phi_0}}}\derivative{S_{{\phi_0}}}{{\phi_0}}}_{R,{\phi_0}}\expval{\derivative{S_{R,{\phi_0}}}{{\phi_0}}}_{R,{\phi_0}} - \expval{e^{-S_{I,{\phi_0}}}\left( \derivative[2]{S_{{\phi_0}}}{{\phi_0}}-\derivative{S_{{\phi_0}}}{{\phi_0}}^2\right) }_{R,{\phi_0}}
\end{align}
We stress that the calculations of these derivatives rely only on a single ensemble.
Ref.~\cite{Alexandru:2018fqp} points out that these derivatives may be simplified and estimated with reliability even in cases with a sign problem.
Practically, they enable iterative procedures for a predicting the optimized shift. When the second derivative is negative we can get a good prediction via Newton-Raphson
\begin{equation}
\phi_{0,i+1} = \phi_{0,i} - \frac{\derivative{}{\phi_0}\expval{e^{-iS_{I,\phi_0}}}_{R,\phi_0,i}}{\derivative{^2}{{\phi_0}^2}\expval{e^{-iS_{I,{\phi_0}}}}_{R,{\phi_0},i}} \ .
\end{equation}
When the second derivative is positive we work with the first derivative to approach the peak. As soon as there are points with opposite first derivatives, their central value usually gets us into the region where the second derivative is negative or at least fairly close to it. 
Additionally we limit the searching region to the interval $[-\tilde{U},+\tilde{U}]$.
In principle, higher order derivatives can also be calculated and used to predict the location of the optimized shift though the statistical errors in these terms grow.
In practice we find that the procedure converges quickly when starting from a region where the sign problem is light enough to calculate at least the first derivative.
However it can fail when this is not the case and it gets stuck when the statistical power is of the same order of magnitude as its uncertainty.
A more quantitative demonstration of this method will be given in \cref{sec:benchmarks}.
A more advanced method might fit all known measurements and derivatives to estimate the location of the peak.

When we are only interested in separate sets of parameters we have to rely on an analytic approximation as the starting point. When we want to scan over one parameter sufficiently finely, we can do so iteratively starting from the previous offset or a rescaled version of it.
In this paper we rescaled it with the fraction of new and previous tangent plane offset.

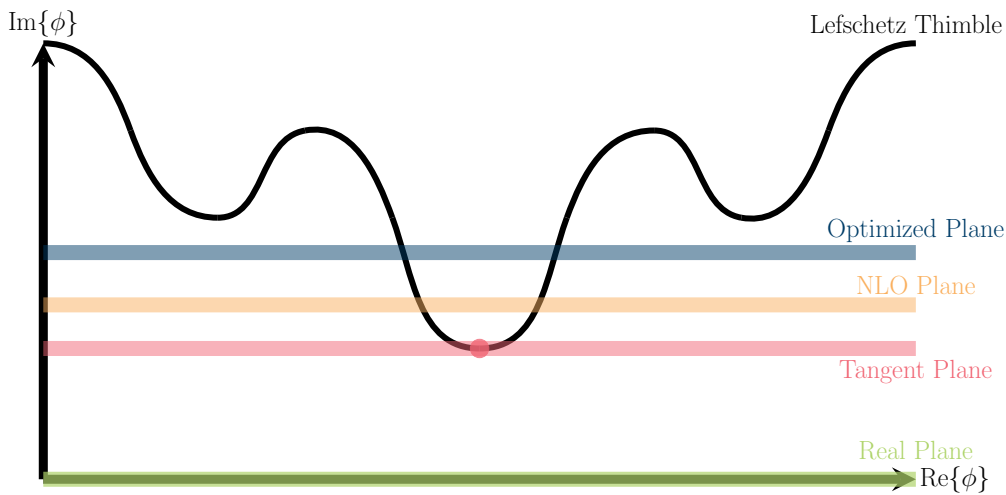
\begin{figure}[h]
	\scalebox{0.4}{\begin{tikzpicture}[node distance = 4em, scale=2.9,font=\Huge]
			\draw[arrow,line width=3mm] (0,0) -- node[at end,yshift=2em] {$\Im{\phi}$} (0,5);
			\draw[arrow,line width=3mm] (0,0) -- node[at end,xshift=4em] {$\Re{\phi}$} (10,0);
			
			% thimble plot
			\draw[black,line width=2mm] (0,5) to[out=0,in=110] (1,4);
			\draw[black,line width=2mm] (1,4) to[out=290,in=180] (2,3);
			\draw[black,line width=2mm] (2,3) to[out=0,in=190] (3,4);
			\draw[black,line width=2mm] (3,4) to[out=10,in=110] (4,3);
			\draw[black,line width=2mm] (4,3) to[out=290,in=180] (5,1.5);
			\draw[black,line width=2mm] (5,1.5) to[out=0,in=-110] (6,3);
			\draw[black,line width=2mm] (6,3) to[out=-290,in=-180] (7,4);
			\draw[black,line width=2mm] (7,4) to[out=-0,in=-190]   (8,3);
			\draw[black,line width=2mm] (8,3) to[out=-10,in=-110]  (9,4);
			\draw[black,line width=2mm] (9,4) to[out=-290,in=-180] node[at end,xshift=-1em,yshift=2em] {Lefschetz Thimble} (10,5);
			
			\draw[fzjgreen,line width=5mm, opacity=0.7] (0,0) to node[at end, above, opacity=1,yshift=1em]{Real Plane} (10,0);
			
			\draw[fzjred,line width=5mm, opacity=0.5] (0,1.5) to node[at end,below, opacity=1.0]{Tangent Plane} (10,1.5);

			\draw[fzjorange,line width=5mm, opacity=0.5] (0,2.0) to node[at end,above, opacity=1.0]{NLO Plane} (10,2.0);
			\draw[fzjred, fill=fzjred, opacity=0.8] (5,1.5) circle(3pt);
			
			\draw[fzjblue1,line width=5mm, opacity=0.5] (0,2.6) to node[at end,above, opacity=1.0]{Optimized Plane} (10,2.6);
	\end{tikzpicture}}
	\caption{A cartoon of the different manifolds referred to throughout this work. The Lefschetz thimbles are drawn to resemble contours of holomorphic flow applied to constant fields. The real plane, tangent plane, NLO estimate (next to leading order correction) and optimized plane show the planar manifolds that we use as our integration regions. The red dot marks the main critical point.
	}
	\label{fig:LefschetzOffsets}
\end{figure}

Figure \ref{fig:LefschetzOffsets} gives a cartoon showing the different offsets in comparison to the Lefschetz thimbles and conveys a geometrically intuitive understanding as to why some planes do better than others.

\begin{figure}[h]
	\centering
	\includegraphics[width=\linewidth]{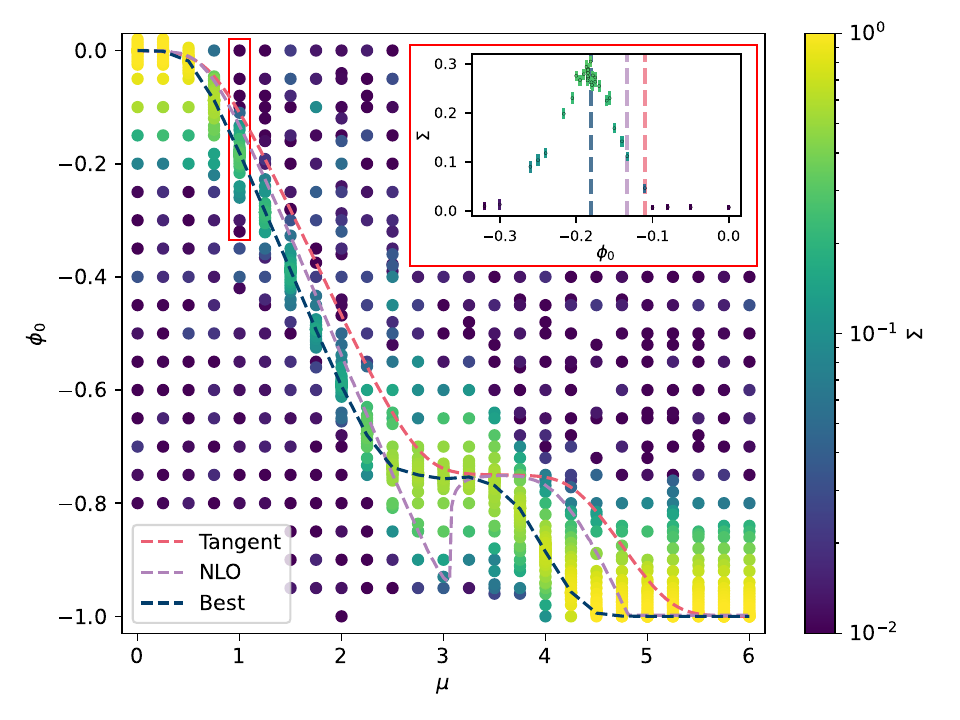}
	\caption{The datapoints in this heatmap show the statistical power of the system as indicated by the colorbar at different chemical potential $\mu$ evaluated with HMC on a plane with an imaginary offset given by the y-axis. The red and purple curve show our analytically determined offsets. The blue curve connects the offsets with the greatest statistical power at each $\mu$. The inset plot shows a slice of the heatmap to visualize the connection to \cref{fig:osp}. The system is the eight site honeycomb lattice with $N_t=16$, $\beta=8$, $U=2$. 
	}
	\label{fig:heatmap}
\end{figure}
\begin{figure}[h]
	\centering
	\includegraphics[width=\linewidth]{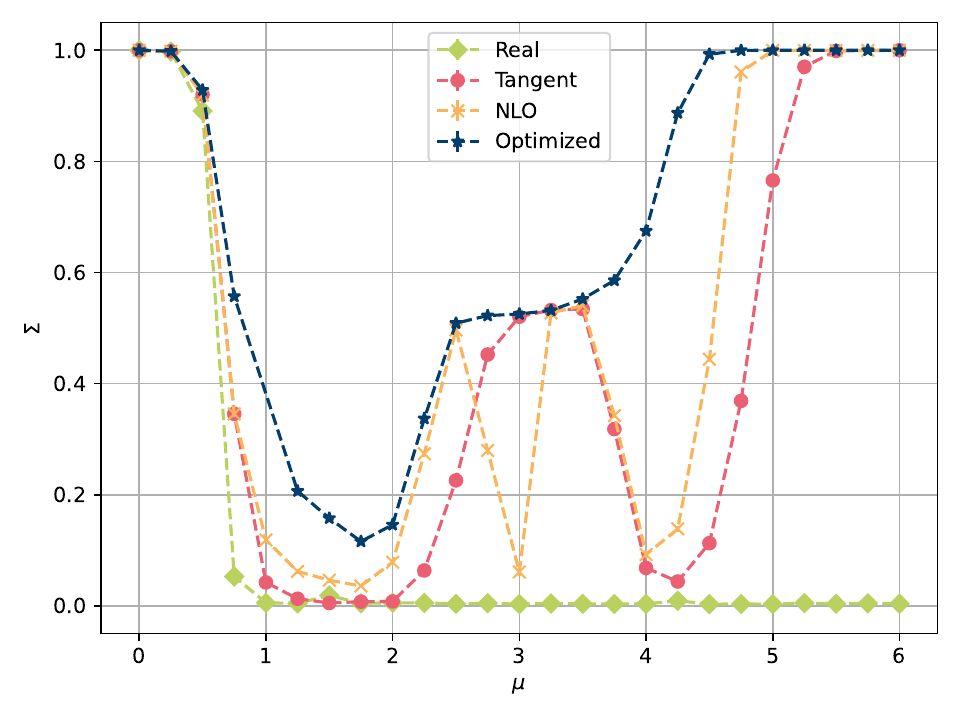}
	\caption{Comparing statistical power of the real plane, tangent plane, NLO correction and optimized shift as a function of $\mu$. The system is an eight site honeycomb lattice with $N_t=16$, $\beta=8$ and $U=2$.}
	\label{fig:sp}
\end{figure}

In \cref{fig:heatmap} we show the statistical power for an interesting range of chemical potentials and imaginary offsets.  We trace contours for the tangent plane, NLO offset, and the best offsets. The best case scenario would be a cheaply determined estimate of the optimized offset for a given $\mu$.
The statistical powers of the real-, tangent- , NLO, and optimized planes are compared as a function of $\mu$ in \cref{fig:sp}.  The key takeaways are that the tangent plane consistently and drastically outperforms the standard algorithm at practically the same cost, and that the sign problem vanishes when the system becomes saturated. Furthermore we see that we can reduce the sign problem for parameters where the tangent plane is insufficient. For system sizes of interest these differences determine whether a system can be calculated (with reasonable resources) or not.
Note that the sign problem also gets worse with increasing $\beta$ and $U$ such that even the optimized plane will eventually fail at finite $\mu$.
This simple method is not suitable to fix the sign problem across the board, it just leads to an efficient expansion of the calculable parameter space, which might or might not include interesting physical phenomena, but definitely enables us to do better zero temperature extrapolations. 
For stronger sign problems we have to rely on manifolds with more parameters, either with simple parametrizations or with neural networks~\cite{Alexandru2016,Wynen2020,Rodekamp2022}.

\subsection{Benchmarks}\label{sec:benchmarks}
\begin{figure}[h]
	\centering
	\begin{subfigure}[t]{0.45\linewidth}
	\includegraphics[width=\linewidth]{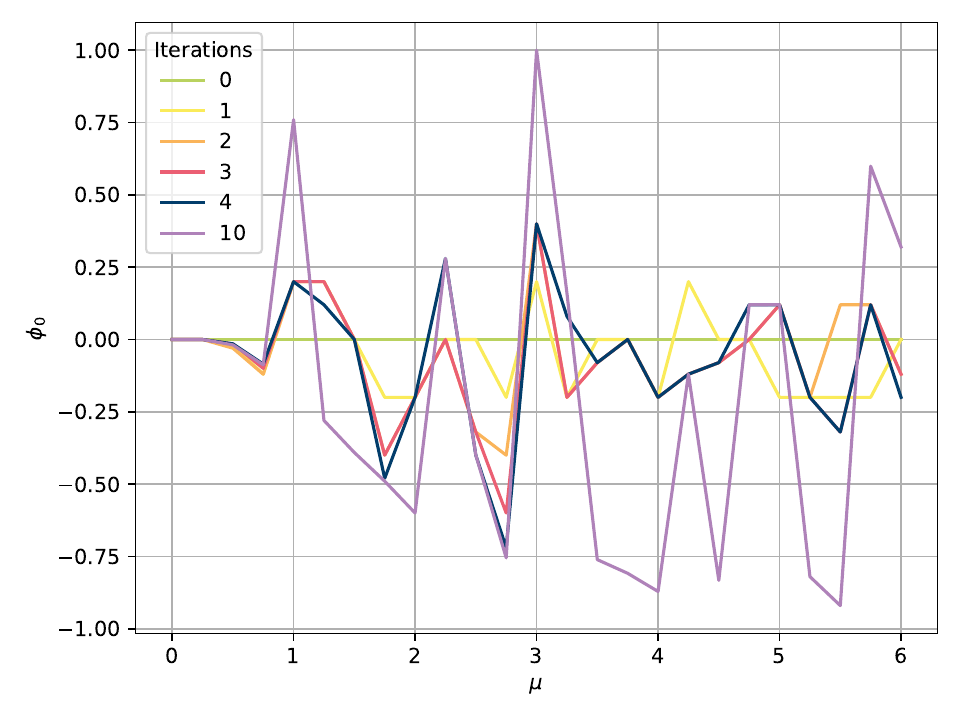}
	\caption{Starting from real plane.}
	\end{subfigure}
	\begin{subfigure}[t]{0.45\linewidth}
		\includegraphics[width=\linewidth]{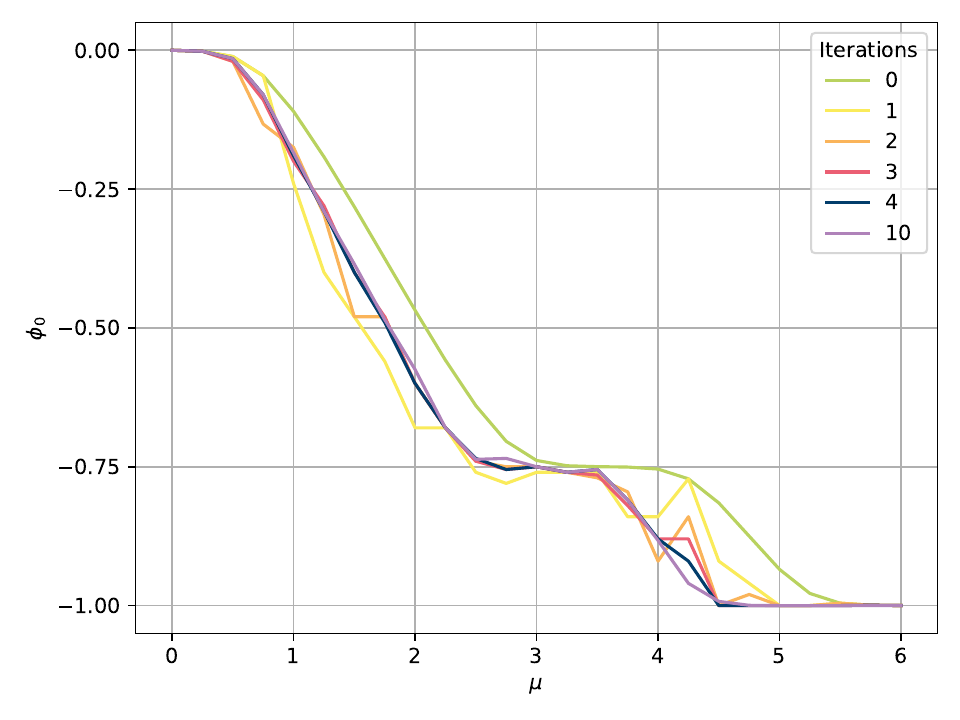}
		\caption{Starting from tangent plane.}
	\end{subfigure}
	\\
	\begin{subfigure}[t]{0.45\linewidth}
		\includegraphics[width=\linewidth]{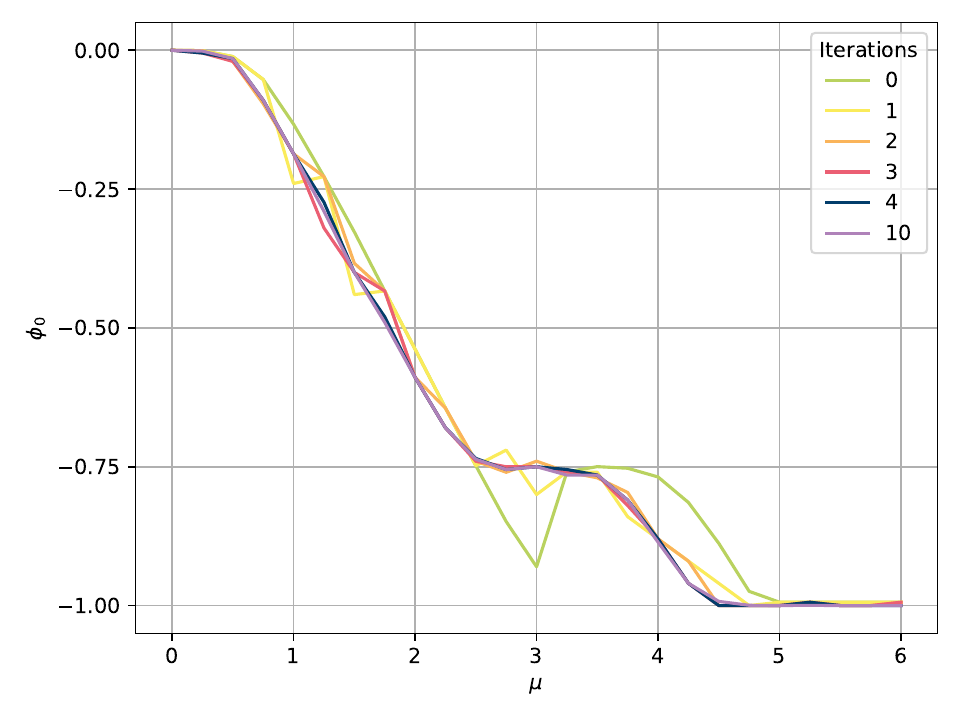}
		\caption{Starting from leading order correction.}
	\end{subfigure}
	\begin{subfigure}[t]{0.45\linewidth}
		\includegraphics[width=\linewidth]{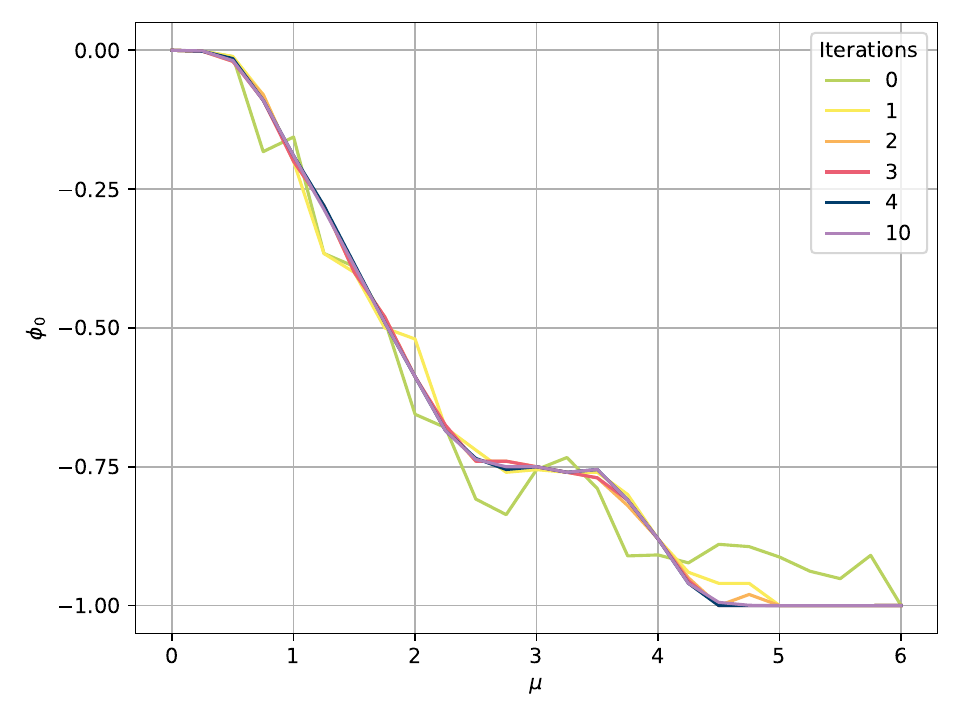}
		\caption{Iteratively starting at rescaled previous result.}
	\end{subfigure}
	\caption{Numerically optimized offset after a given number of iterations. Iteration 0 marks the starting offsets. This plot shows the convergence to the optimal offset and the value of a good starting guess. The iterative method started from $\mu=0$ and $\mu=6$, meeting in the middle.}
	\label{fig:offsetEvolution}
\end{figure}

\begin{figure}[h]
	\centering
	\begin{subfigure}[t]{0.45\linewidth}
		\includegraphics[width=\linewidth]{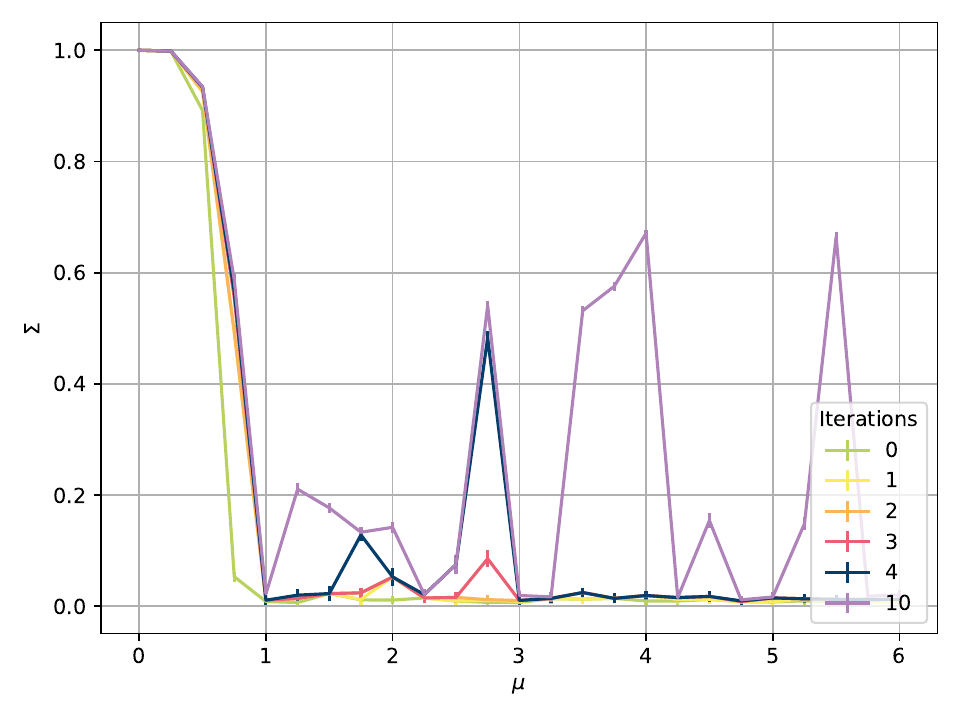}
		\caption{Starting from real plane.}
	\end{subfigure}
	\begin{subfigure}[t]{0.45\linewidth}
		\includegraphics[width=\linewidth]{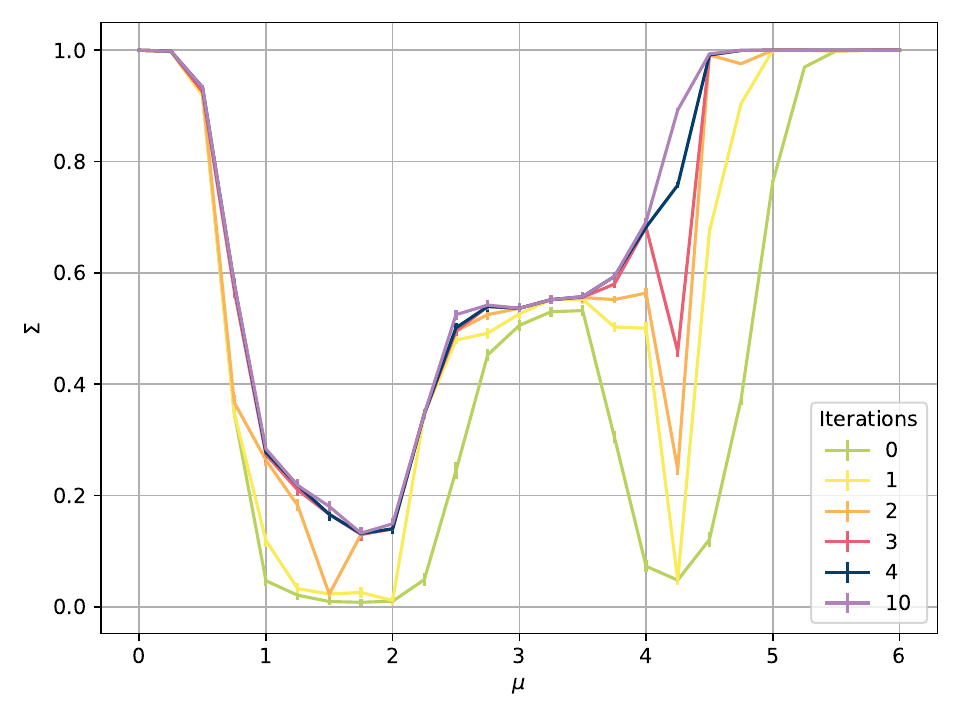}
		\caption{Starting from tangent plane.}
	\end{subfigure}
	\\
	\begin{subfigure}[t]{0.45\linewidth}
		\includegraphics[width=\linewidth]{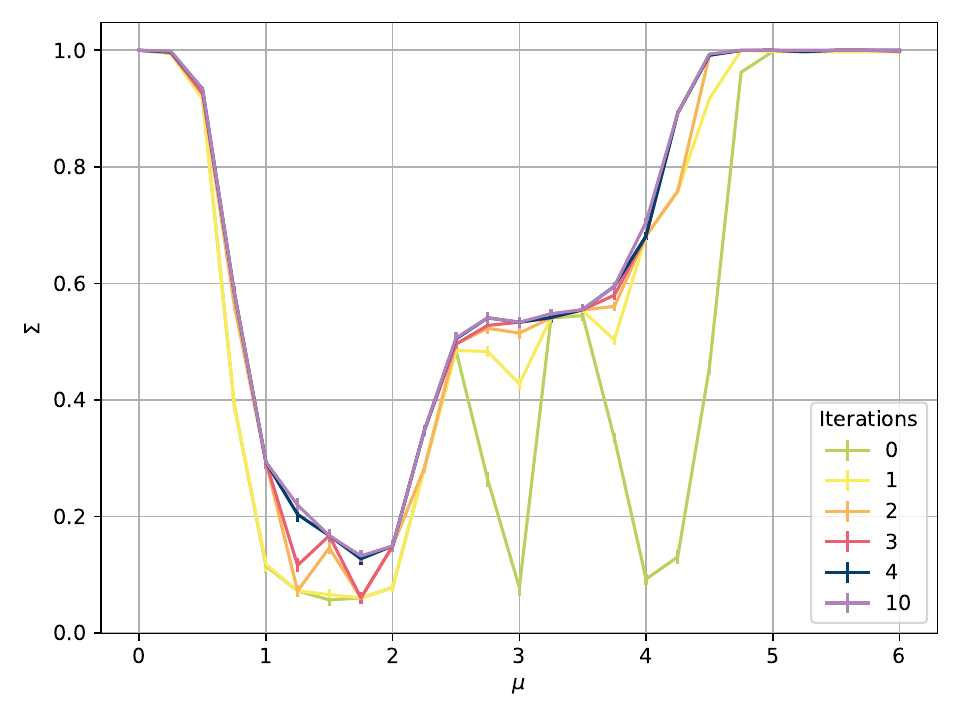}
		\caption{Starting from leading order correction.}
	\end{subfigure}
	\begin{subfigure}[t]{0.45\linewidth}
		\includegraphics[width=\linewidth]{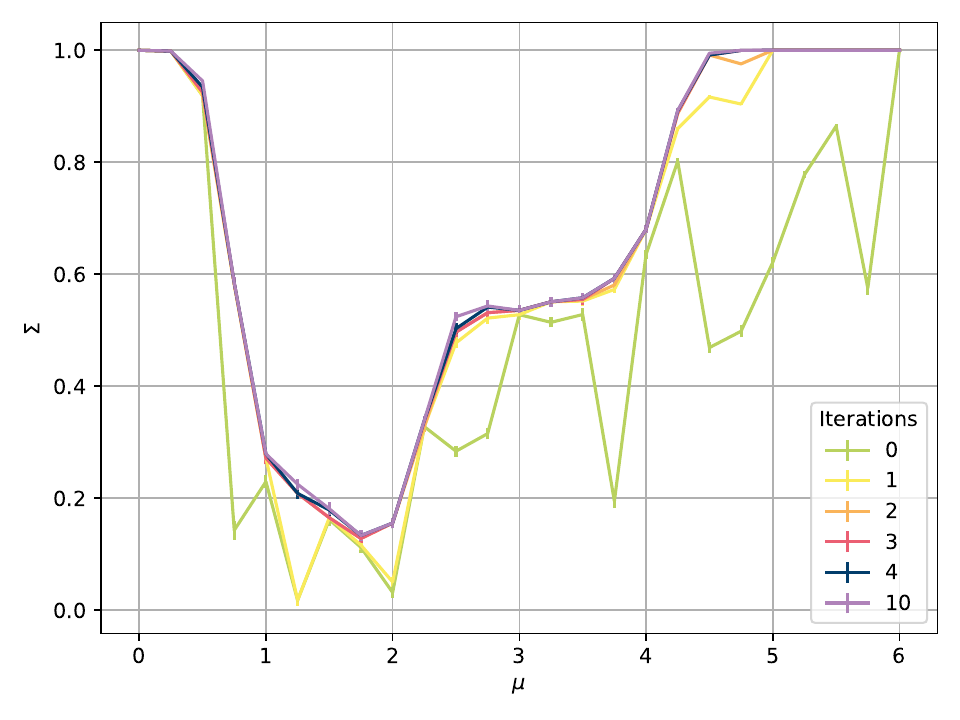}
		\caption{Iteratively starting at rescaled previous result.}
	\end{subfigure}
	\caption{Statistical power at numerically determined optimal offset, after a given number of optimization steps. For each iteration we used $\num{10000}$ HMC steps (+tuning).}
	\label{fig:spEvolution}
\end{figure}

In this section we demonstrate the convergence of our numerical optimization algorithm and present the resulting increase of the statistical power. The examples refer to the eight site honeycomb lattice with $N_t=16$, $\beta=8$ and $U=2$. Figure \ref{fig:offsetEvolution} showcases the convergence with iterations from different starting points. While they would all converge to the same offsets eventually, we observe that starting in a region with unclear first derivative, that is dominated by statistical noise, turns the algorithm into a random walk, which can be observed in the real plane example. This further highlights the importance of having good analytic starting points. 
Figure \ref{fig:spEvolution} shows the significant improvements in statistical power that can be achieved by just a few iterations. We observe in both figures that most runs starting from a reasonable guess converge and roughly agree with each other after just 3 iterations. Here the improvements from tangent plane to leading order correction to iterative starting points can be best observed by comparing the second iterations with each other and and in comparison to the converged result.

%% file: section/results.tex
%!TEX root = ../master.tex
\section{Results}\label{sec:results}

\begin{figure}[h]
	\centering
	\begin{subfigure}[t]{0.45\linewidth}
		\includegraphics[width=\linewidth]{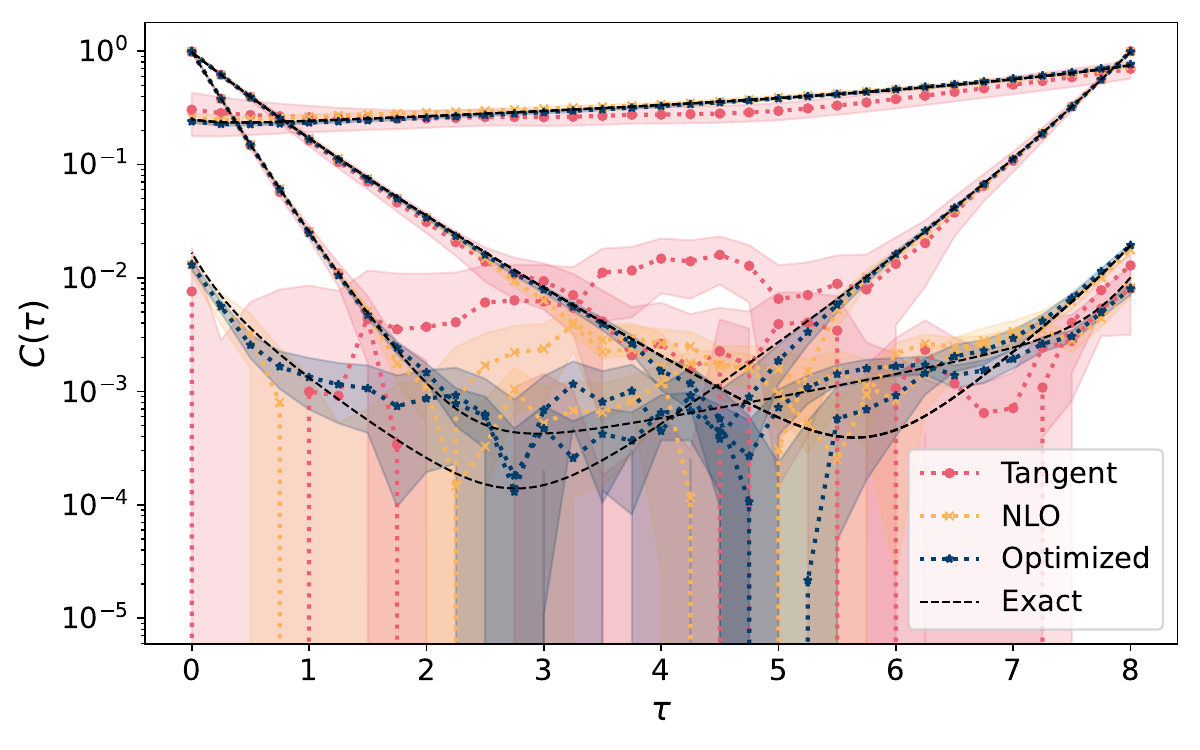}
		\caption{Eight site honeycomb lattice at $\beta=8$.}
		\label{fig:8C}
	\end{subfigure}
	\begin{subfigure}[t]{0.45\linewidth}
		\includegraphics[width=\linewidth]{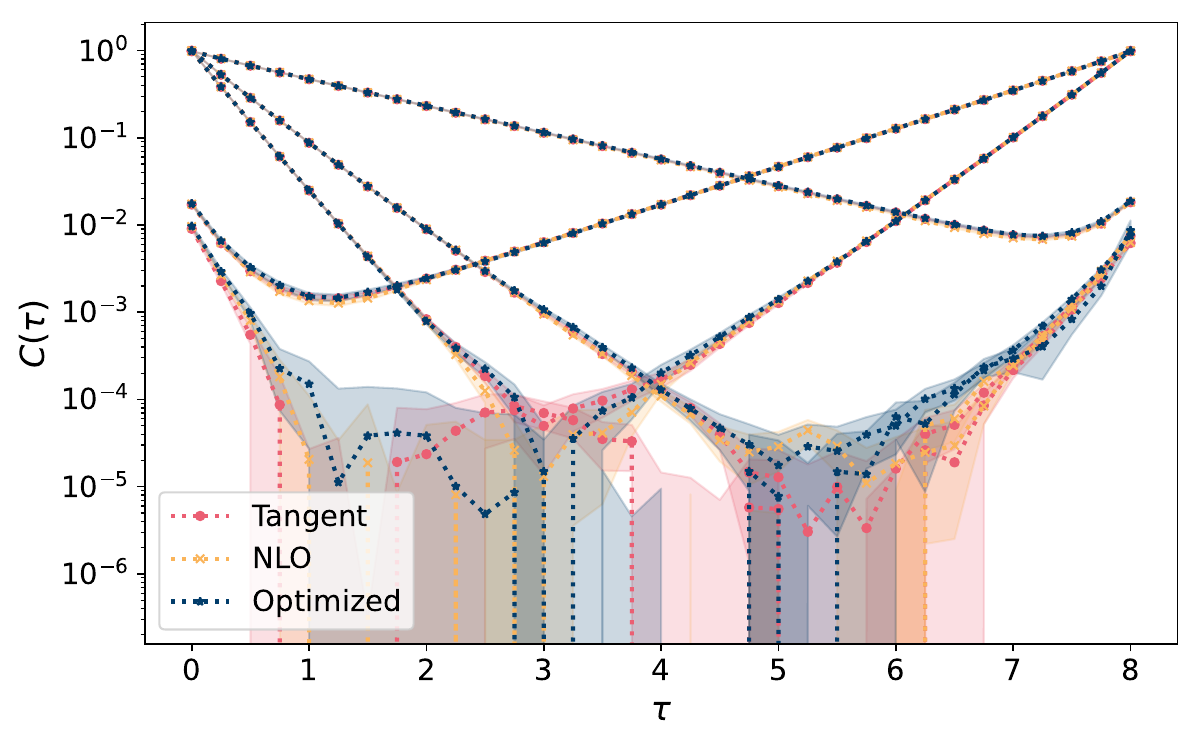}
		\caption{18 site honeycomb lattice at $\beta=8$.}
		\label{fig:18C}
	\end{subfigure}
	\\
	\begin{subfigure}[t]{0.45\linewidth}
		\includegraphics[width=\linewidth]{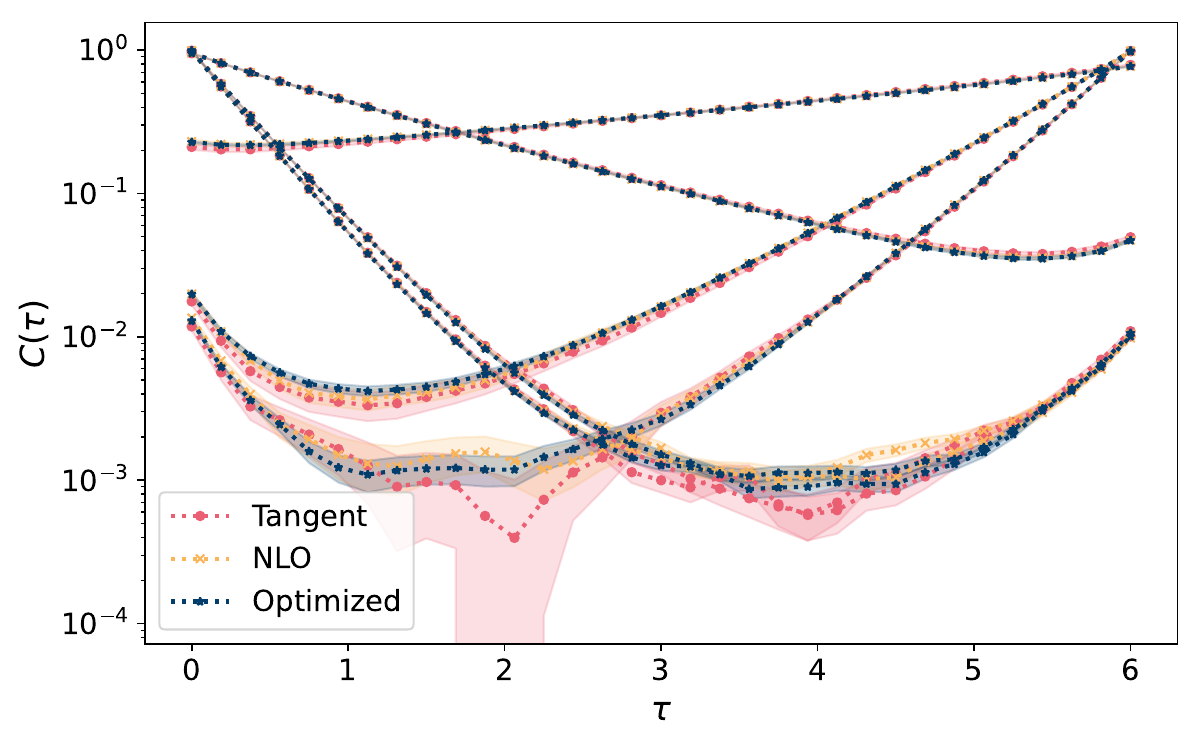}
		\caption{$C_{20}$ at $\beta=6$.}
		\label{fig:20C}
	\end{subfigure}
	\begin{subfigure}[t]{0.45\linewidth}
		\includegraphics[width=\linewidth]{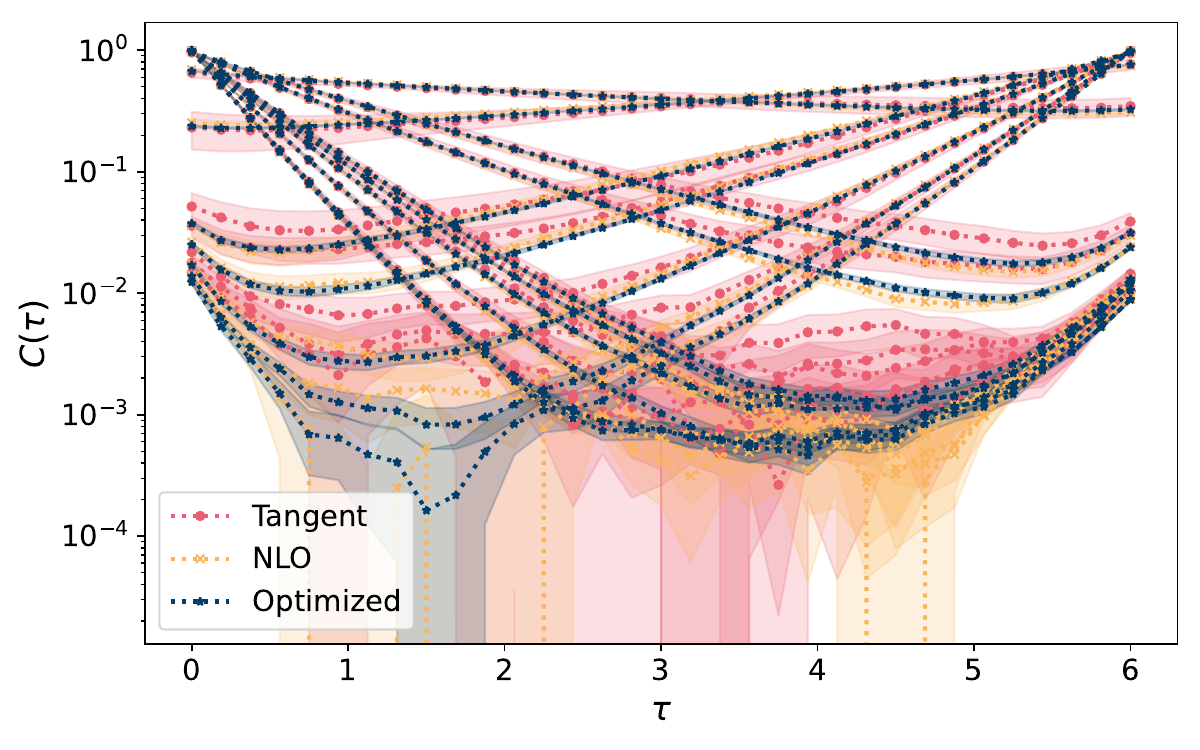}
		\caption{$C_{60}$ at $\beta=6$.}
		\label{fig:60C}
	\end{subfigure}
	\caption{Single particle correlation function for different lattices and calculated with different methods. All systems were evaluated with $N_t=32$ at $U=2$ and $\mu=1$. Each data point was calculated from a markov chain with $50,000$ HMC configurations, where we measured on each $10^{\text{th}}$ to reduce autocorrelation, furthermore we averaged over correlators guaranteed to be equal by symmetry. The exact solution in \cref{fig:8C} was determined by exact diagonalization in the temporal continuum limit.}
	\label{fig:correlators}
\end{figure}

In this section we provide physical observables determined by HMC with the introduced modifications. The observables of our choice are the single particle correlation functions 
\begin{equation}\label{eq:spc}
	C_k(\tau) = \expval{a_k^{}(\tau)a_k^\dagger(0)}
\end{equation}
where $a$ and $a\adjoint$ are particle ladder operators and $k$ labels operators in definite the irreducible representations of the lattice automorphism group.
In the honeycomb case $k$ labels operators with definite momentum; for the fullerenes $k$ labels representations of the icosahedral symmetry group.
We also sum the local charges \eqref{eq:H and q} to measure the global charge
\begin{equation}\label{eq:charge}
	\expval{Q} = \expval{\sum_x q_x } = N_x - 2\sum_k C_k(\tau=0)\ .
\end{equation}
To establish trust in the physical correctness of the algorithm we compare with the eight site honeycomb lattice for which we have exact results from direct diagonalization.
We find that the uncertainties of the standard real plane HMC are much greater than the uncertainties of our more advanced methods. Especially for the eight site lattice  the real plane results do not agree with the exact solution while the calculations with an imaginary offset match it very well. Above all the optimized offset resembles the exact solution with great precision. This plus the agreement with the NLO, and often with the tangent plane as well, gives us confidence in the results of the larger systems.
We present the correlation functions resulting from our methods in \cref{fig:correlators}.

\begin{figure}[h]
	\centering
	\begin{subfigure}[t]{0.45\linewidth}
		\includegraphics[width=\linewidth]{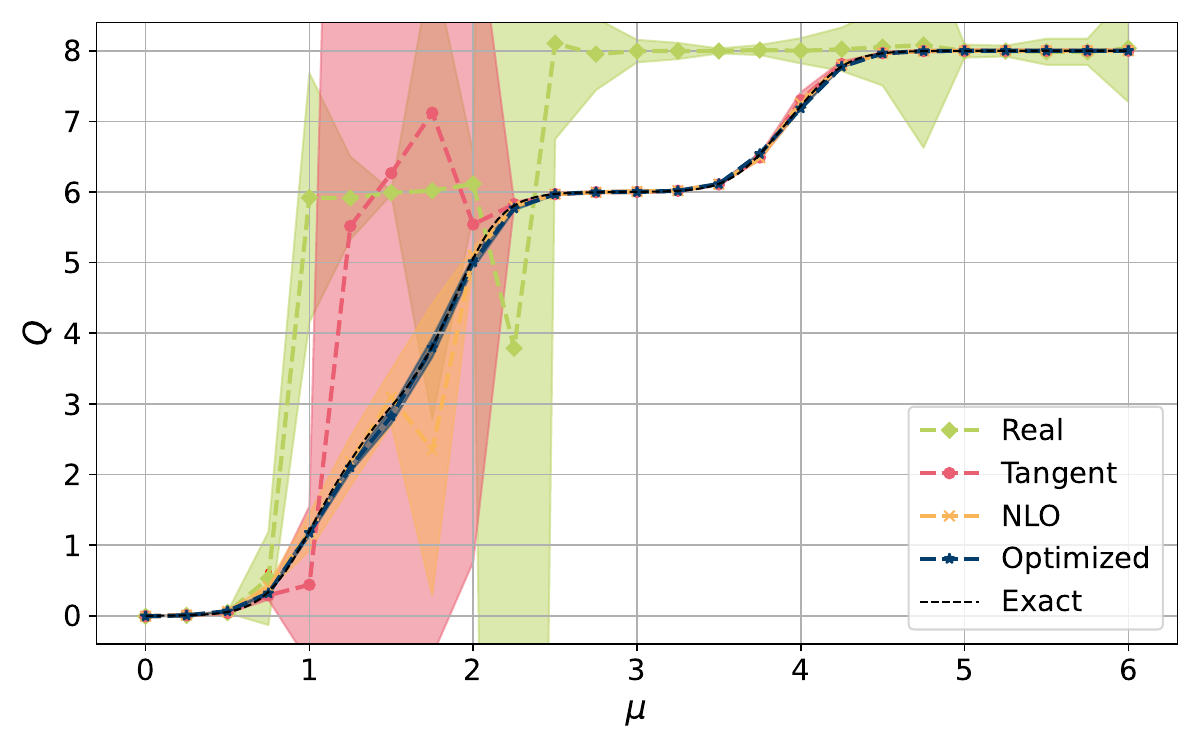}
		\caption{Eight site honeycomb lattice at $\beta=8$.}
		\label{fig:8Q}
	\end{subfigure}
	\begin{subfigure}[t]{0.45\linewidth}
		\includegraphics[width=\linewidth]{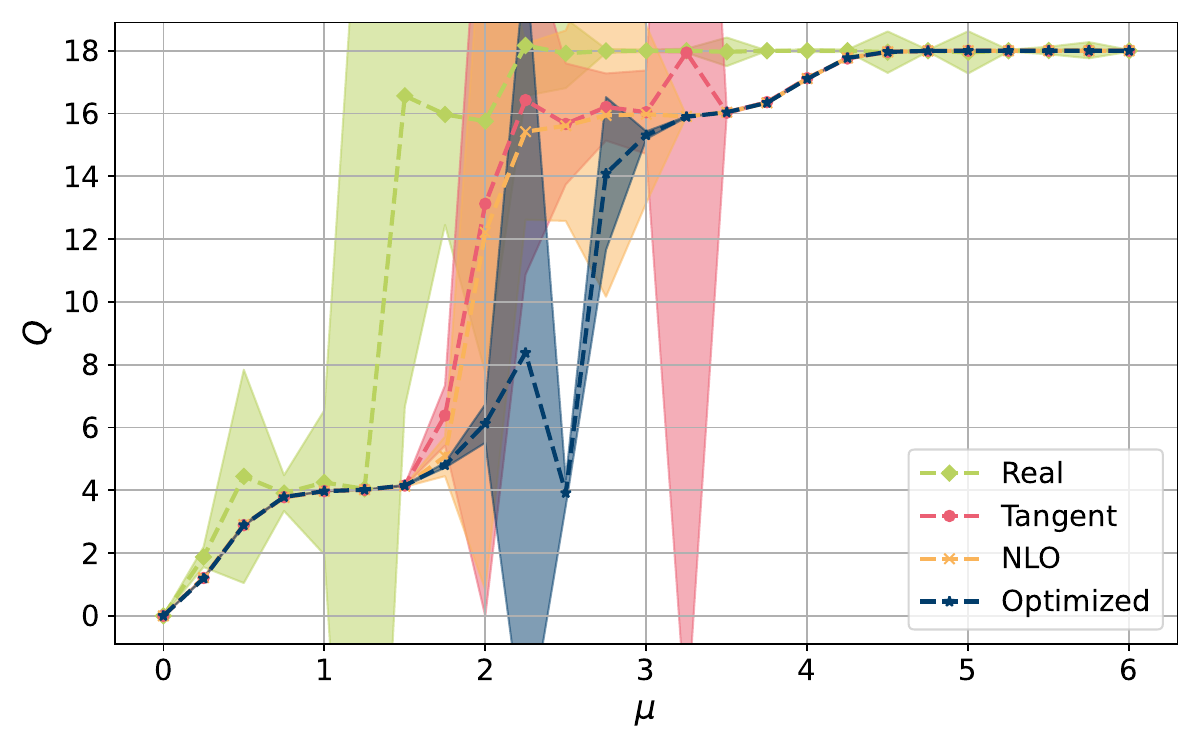}
		\caption{18 site honeycomb lattice at $\beta=8$.}
		\label{fig:18Q}
	\end{subfigure}
	\\
	\begin{subfigure}[t]{0.45\linewidth}
		\includegraphics[width=\linewidth]{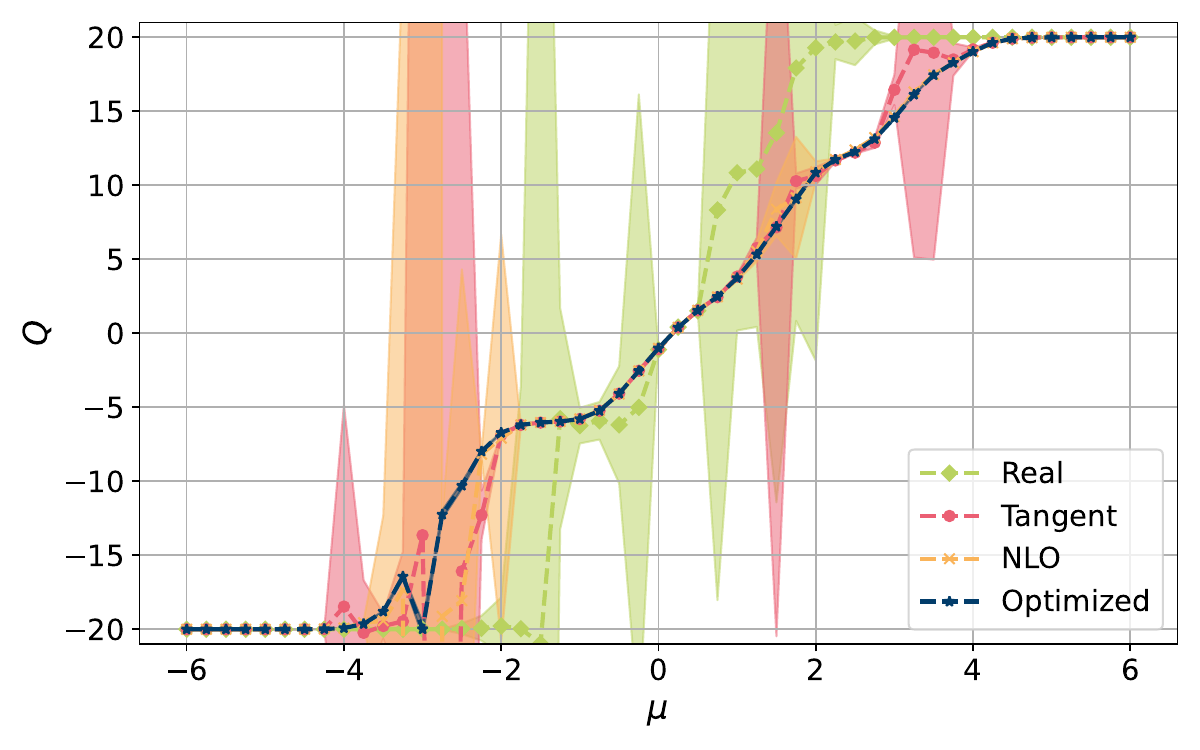}
		\caption{$C_{20}$ at $\beta=6$.}
		\label{fig:20Q}
	\end{subfigure}
	\begin{subfigure}[t]{0.45\linewidth}
		\includegraphics[width=\linewidth]{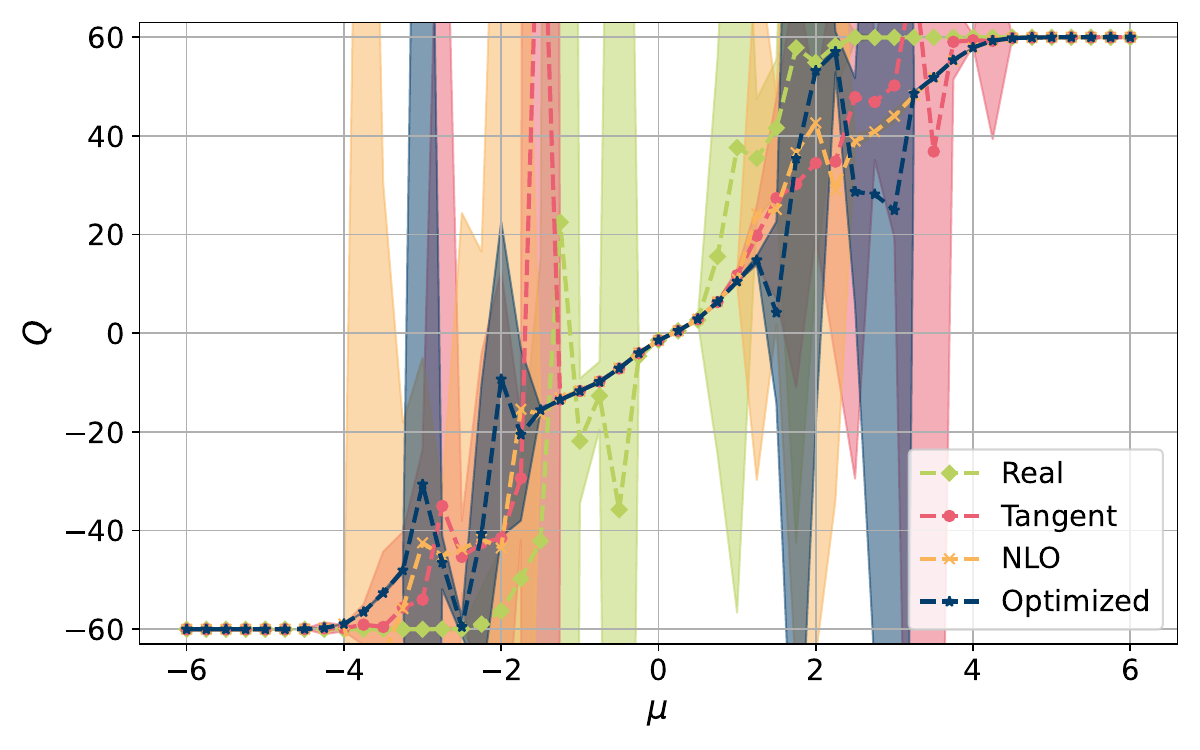}
		\caption{$C_{60}$ at $\beta=6$.}
		\label{fig:60Q}
	\end{subfigure}
	\caption{Charge expectation value at varying $\mu$ for different lattices and calculated with different methods. All systems have $N_t=16$ and $U=2$. Each data point was calculated from a markov chain with $\num{50000}$ HMC configurations, where we measured on each $10^{\text{th}}$ to reduce autocorrelation. The exact solution in \cref{fig:8Q} was determined by exact diagonalization with $N_t=16$ discretization.}
	\label{fig:charge}
\end{figure}

\Figref{charge} shows that for the same number of configurations the quality of the measured observables seems to match the expected outcome from comparing the statistical powers, which can be found in \cref{fig:sp} and \cref{fig:muSPa}. 
We see larger statistical fluctuations with worse statistical power; the optimized method consistently performs best.
\Figref{8Q} shows that most of the numerical results estimate the charge correctly \eqref{eq:charge} according to the exact results. The real plane HMC underestimates its error systematically for large ranges of $\mu$.
Still, the optimized offset resembles the exact result best; the tangent plane and NLO arguably offer comparable uncertainties for many parameter choices.
\Figref{60Q} and \cref{fig:60SP} show that our method has limits and not every sign problem can be conquered with a simple constant offset.
Also in certain areas of the other charge plots, we see that the optimization routine could fail when the sign problem is very strong, causing a worse result than the NLO.
An interesting observation is that where the charge flattens the statistical power of the tangent and NLO planes peak.

%% file: section/conclusion.tex
%!TEX root = ../master.tex
\section{Conclusion}\label{sec:conc}

Our results clearly show that the simple introduction of an imaginary shift of the integration contour can greatly impact the severity of the sign problem in quantum field theory with practically no additional computation cost or human effort required. We provide two analytic expressions for such offsets for the Hubbard model. Further a careful tuning of these offsets while coming at a small cost can lead to even better outcomes, especially when a range of parameters is to be explored. The reduction of the sign problem depends on the system and the physical parameters, but can potentially make a difference of orders of magnitude which can be seen as an increase in measurement precision at a fixed sample size or saving of computational resources on the way to achieve a certain desired precision. Even though this does not eliminate the sign problem entirely, it extends the parameter space that is explorable within our naturally limited resources and allows us to perform higher quality extrapolations.

In addition to the analysis of the method itself we provided observables to condensed matter systems which could not be calculated before with lattice stochastic methods, going as far as the $C_{60}$ buckyball a stable synthesizable carbon nanosystem.
Further we found that in the limit of infinite chemical potential, which in case of the Hubbard model refers to a completely filled/empty lattice, the sign problem vanishes at a certain offset stating the question if this behavior could be seen in other theories as well.
Our numerical optimization could lend itself to unsupervised learning, driving towards maximizing the statistical power and minimizing its derivative \eqref{eq:dSP}.

There are some open questions remaining in regards to combining optimized offset with neural networks, that we will address in the future. Is the optimized offset also the best starting point for generating training data? Does the uplift of neural networks and optimized shift over the tangent plane correlate? Further we plan to investigate the $C_{60}$ lattice in more detail, as our methods have opened the door to high quality measurements on these large systems that are out of reach for exact diagonalization as well as HMC without sign problem optimization.
We are developing a library for these nanosystems calculations with the intention of making it publicly available so everyone can try out our methods on the theory or model they are interested in.

%% file: section/appendix.tex
%!TEX root = ../master.tex
\newpage
\section{Statistical Power}
\begin{figure}[h]
	\centering
	\begin{subfigure}[t]{0.45\linewidth}
		\includegraphics[width=\linewidth]{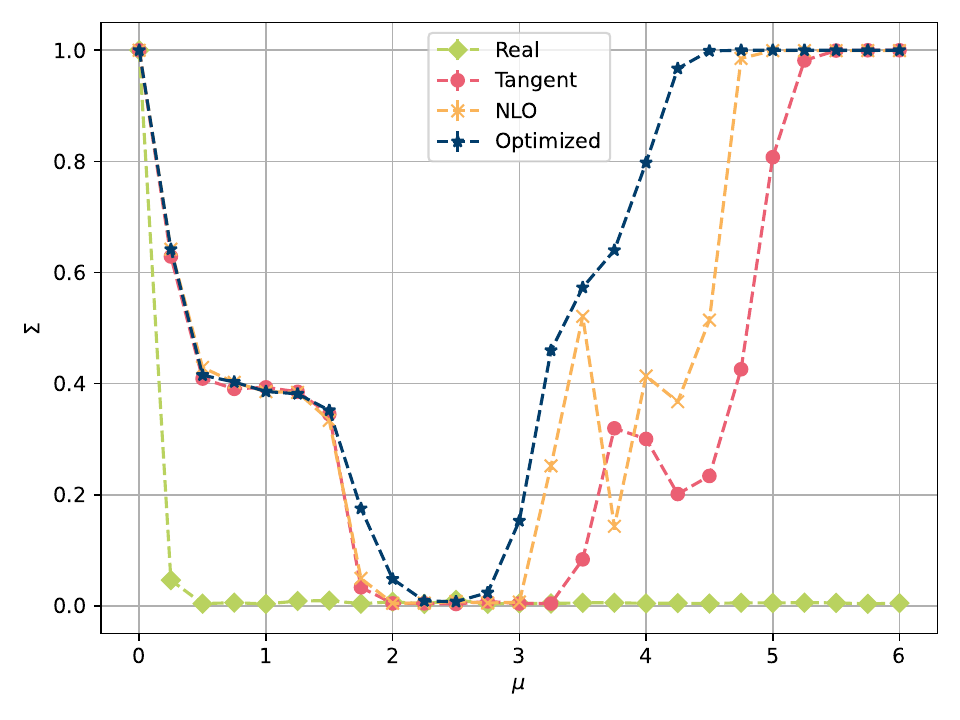}
		\caption{18 site honeycomb lattice with $\beta=8$.}
	\end{subfigure}
	\begin{subfigure}[t]{0.45\linewidth}
		\includegraphics[width=\linewidth]{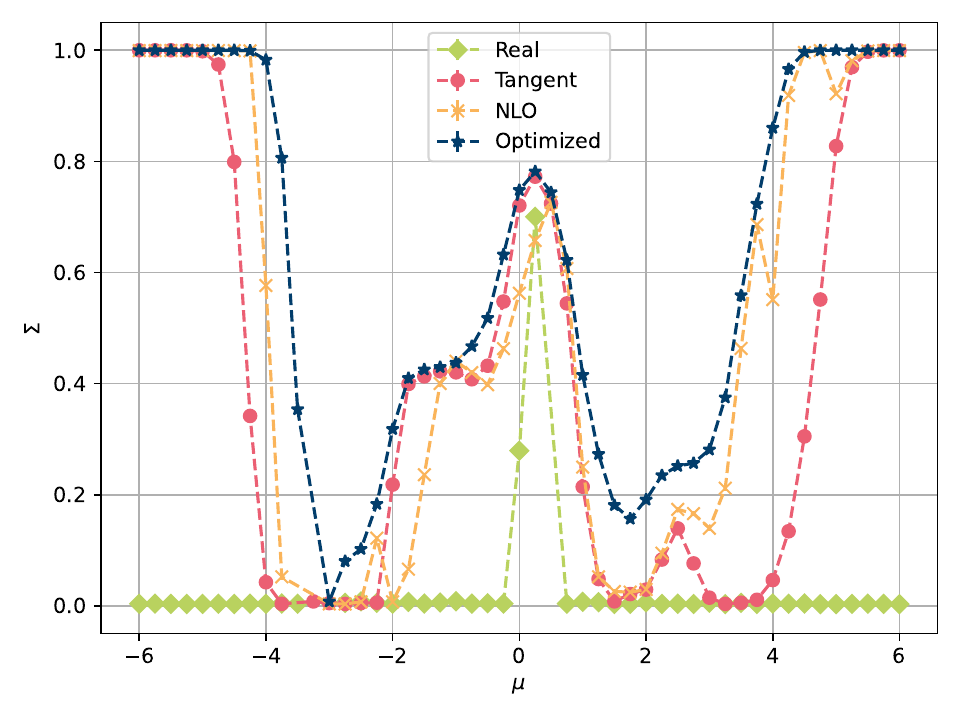}
		\caption{$C_{20}$ lattice with $\beta=6$.}
	\end{subfigure}
	\\
	\begin{subfigure}[t]{0.45\linewidth}
		\includegraphics[width=\linewidth]{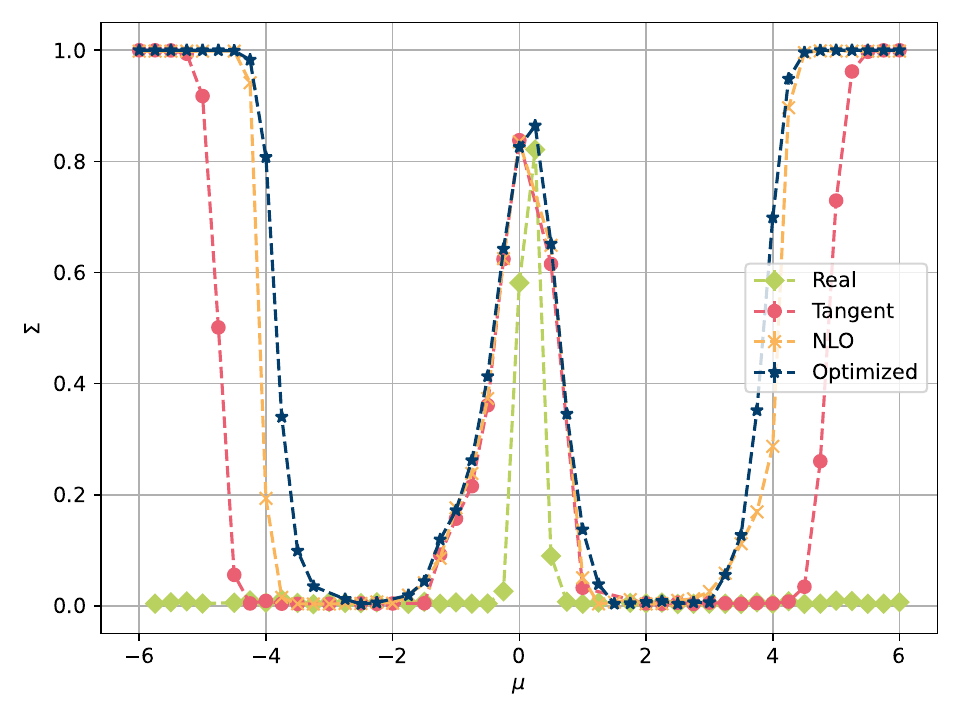}
		\caption{$C_{60}$ lattice with $\beta=6$.}
		\label{fig:60SP}
	\end{subfigure}
	\caption{Statistical power for different lattices over a range of $\mu$. For all systems $U=2$ and $N_t=16$.}
	\label{fig:muSPa}
\end{figure}

In \cref{fig:muSPa} we show the behavior of the statistical power over the interesting range of $\mu$ for the remaining lattices, where we again compare the different offset methods. We also see in the plot that for non-bipartite lattices introducing a small chemical potential can relieve the sign problem.

\section{Detailed derivation of tangent plane}\label{sec:appendix tangent plane}

Starting from the action \eqref{eq:action} we provide a detailed derivation the equation that determines the tangent plane \eqref{eq:transcendental}.
As in the main text \eqref{eq:MU} let
\begin{align}
	\FF_\pm = \FF[\pm\phi, \pm K, \pm \mu] &= \prod_{\tau=0}^{N_t-1} f^\pm_\tau
	&
	f^\pm_\tau = 
	\left[e^{\pm\tilde{K}}\right] \left[e^{\pm(-i\phi_{x,\tau}+\tilde{\mu})}\right]
	\label{eq:F}
\end{align}
where each term in $[$square brackets$]$ is a space-by-space matrix and the product is from right ($\tau=0$) to left $(\tau=N_t-1)$.
Then, using the Schur complement
\begin{align}\label{eq:schur}
	\det M[\phi, K, \mu] = \det\left( \one + \FF[\phi, K, \mu] \right)
\end{align}
and likewise for $M[-\phi,-K,-\mu]$.  Since the inverses exist for any configuration with finite action
\begin{align}
	\label{eq:dS}
	\frac{\partial S}{\partial \phi_{x,t}} &= \oneover{\tilde{U}} \phi_{x,t} - \tr{(\one+\FF_+)\inverse \partial_{\phi_{x,t}} \FF_+} - \tr{(\one+\FF_-)\inverse \partial_{\phi_{x,t}} \FF_-}
\end{align}
where the two traces correspond to the particle and hole fermion matrices \eqref{eq:MU}.

Each auxiliary field only appears once in any given $\FF$, inside one entry of a diagonal matrix.
So, differentiating $\FF_{\pm}$ inserts $\mp i \PP_x$ where $\PP_x$ projects to site $x$,
\begin{align}
	\label{eq:dF}
	\partial_{\phi_{x,t}} \FF_{\pm} &= 
	\left[\prod_{\tau=t}^{N_t-1} f^\pm_\tau \right] \left[ \mp i \PP_x\right] \left[\prod_{\tau=0}^{t-1} f^\pm_\tau \right]
\end{align}
where the products go from right to left.
Cycling the traces so that the projector is rightmost gives
\begin{align}
	\label{eq:dS2}
	\frac{\partial S}{\partial \phi_{x,t}} &= \oneover{\tilde{U}} \phi_{x,t} + i \sum_s s \tr{
		\left[\prod_{\tau=0}^{t-1} f^\pm_\tau \right]
		(\one+\FF_s)\inverse
		\left[\prod_{\tau=t}^{N_t-1} f^\pm_\tau \right]
		\PP_x
	}
\end{align}
where the sum over $s$ runs over $\pm 1$.

Plugging in the imaginary spacetime constant $\phi = i \phi_0$ means the auxiliary field factors are proportional to the identity matrix,
\begin{align}
	f^\pm_\tau = \exp\left(\pm(\tilde{K} + \phi_0 + \tilde{\mu})\right)
	\label{eq:time independent f}
\end{align}
independent of timeslice $\tau$.
Since $\phi_0$ is space-independent we can sum on $x$ and use the completeness $\sum_x \PP_x = \one$.
Since $N_t \tilde{K} = \beta K$ and $N_t \tilde{\mu} = \beta \mu$,
\begin{align}
	0 =
	\oneover{\tilde{U}} N_x \phi_0 + \sum_{s\in{\pm1}} s \tr{
		\frac{ e^{s(\beta K + N_t \phi_0 + \beta \mu)} }{ 1+ e^{s(\beta K + N_t \phi_0 + \beta \mu)} }
	}
\end{align}
and evaluating this relationship in the eigenbasis of $K$ yields the tangent plane relation \eqref{eq:transcendental}.

\section{Detailed derivation of NLO}\label{sec:appendix NLO}

To find the NLO constant imaginary offset we need to minimize the effective action \eqref{eq:second_order_action} which requires computing the Hessian
\begin{align}
	\HH_{x't',xt}
	=
	\left(\partial_{x't'}\partial_{xt^{}}S[\phi]\right)\vline_{\phi=i\phi_1} \ .
\end{align}
We start from the general single derivative \eqref{eq:dS} and differentiate again.
Without loss of generality we assume $t'\geq t$,
\begin{align}
	\partial_{x't'}\partial_{xt^{}}S[\phi]
	=
	\oneover{\tilde{U}} \delta_{x'x}\delta_{t't}
	& - \sum_s \tr{
		(\one+\FF_s)\inverse \partial_{\phi_{x',t'}} \partial_{\phi_{x,t}} \FF_s
	-	(\one+\FF_s)\inverse \left(\partial_{\phi_{x',t'}} \FF_s\right) (\one+\FF_s)\inverse \left(\partial_{\phi_{x,t}} \FF_s\right)
	}
\end{align}
The second derivative of $\FF$ is much like the first \eqref{eq:dF} but with a second projector $\PP_{x'}$ inserted at time $t'$; the $(\mp i)^2=-1$ regardless of sign choice.
\begin{align}
	\partial_{x't'}\partial_{xt} \FF_{s}
	&=
	-
	\left[\prod_{\tau=t'}^{N_t-1} f^s_\tau\right]
	\PP_{x'}
	\left[\prod_{\tau=t}^{t'-1} f^s_\tau\right]
	\PP_{x}
	\left[\prod_{\tau=0}^{t-1} f^s_\tau \right].
\end{align}
In fact, this looks much like the two-inverse term, though that term has an inverse between the projectors,
\begin{align}
	\left(\partial_{\phi_{x',t'}} \FF_s\right) (\one+\FF_s)\inverse \left(\partial_{\phi_{x,t}} \FF_s\right)
	&=
	-
	\left[\prod_{\tau=t'}^{N_t-1} f^s_\tau\right]
	\PP_{x'}
	\left[\prod_{\tau=0}^{t'-1} f^s_\tau\right]
	\left[\one+\FF_s\right]\inverse
	\left[\prod_{\tau=t}^{N_t-1} f^s_\tau\right]
	\PP_{x}
	\left[\prod_{\tau=0}^{t-1} f^s_\tau \right].
\end{align}
Cycling the trace so that $\PP_x$ is rightmost and consolidating like factors gives
\begin{align}
	\partial_{x't'}\partial_{xt^{}}S[\phi]
	=
	\oneover{\tilde{U}} \delta_{x'x}\delta_{t't}
	& + \sum_s \tr{
		\left[\prod_{\tau=0}^{t-1} f^s_\tau \right]
		(\one+\FF_s)\inverse 
		\left[\prod_{\tau=t'}^{N_t-1} f^s_\tau\right]
		\PP_{x'}
		\left[\prod_{\tau=t}^{t'-1} f^s_\tau\right]
		\left[
			\one
		-	\left[\prod_{\tau=0}^{t-1} f^s_\tau\right]
			\left[\one+\FF_s\right]\inverse
			\left[\prod_{\tau=t}^{N_t-1} f^s_\tau\right]
		\right]
		\PP_{x}
	}.
\end{align}
Making repeated use of $C\inverse B\inverse A\inverse = (ABC)\inverse$ we can re-express the term in the sum
\begin{align}
	\left[\prod_{\tau=0}^{t-1} f^s_\tau\right]
		\left[\one+\FF_s\right]\inverse
		\left[\prod_{\tau=t}^{N_t-1} f^s_\tau\right]
	&=
	\left[\prod_{\tau=t-1}^{0} (f^s_\tau)\inverse\right]\inverse
		\left[\one+\prod_{\tau=0}^{N_t-1} f^s_\tau \right]\inverse
		\left[\prod_{\tau=N_t-1}^{t} (f^s_\tau)\inverse\right]\inverse
	\nonumber\\
	&=
	\left[
			\left[\prod_{\tau=N_t-1}^{t} (f^s_\tau)\inverse\right]
			\left[\one+\prod_{\tau=0}^{N_t-1} f^s_\tau \right]
			\left[\prod_{\tau=t-1}^{0} (f^s_\tau)\inverse\right]
		\right]\inverse
	\nonumber\\
	&=
		\left[
		\left[\prod_{\tau=t-1}^{t} \left(f^s_\tau\right)\inverse \right]
			+\one
		\right]\inverse
	\label{eq:reverse}
\end{align}
where the products of $f\inverse$s count down from right to left and wrap from $0$ to $N_t-1$.
Inserting convenient products equal to the identity to use the result \eqref{eq:reverse} twice gives
\begin{align}
	\partial_{x't'}\partial_{xt^{}}S[\phi]
	=&
	\oneover{\tilde{U}} \delta_{x'x}\delta_{t't}
	\\\nonumber
	& + \sum_s \tr{
		\left[
		\one + \prod_{\tau=t-1}^{t} \left(f^s_\tau\right)\inverse
		\right]\inverse
	\left[\prod_{\tau=t'-1}^{t} (f^s_\tau)\inverse\right]
		\PP_{x'}
		\left[\prod_{\tau=t}^{t'-1} f^s_\tau \right]
		\left[\prod_{\tau=t-1}^{t} \left(f^s_\tau\right)\inverse \right]
		\left[
		\one + \prod_{\tau=t-1}^{t} \left(f^s_\tau\right)\inverse
		\right]\inverse
		\PP_{x}
	}.
\end{align}
Casting the inverse factors after $\PP_{x'}$ into the denominator we arrive at
\begin{align}
	\partial_{x't'}\partial_{xt^{}}S[\phi]
	=
	\oneover{\tilde{U}} \delta_{x'x}\delta_{t't}
	& + \sum_s \tr{
		\left[
		\one + \prod_{\tau=t-1}^{t} \left(f^s_\tau\right)\inverse
		\right]\inverse
		\left[\prod_{\tau=t'-1}^{t} (f^s_\tau)\inverse\right]
		\PP_{x'}
		\left[\prod_{\tau=t}^{t'-1} f^s_\tau \right]
		\left[
		\one + \prod_{\tau=t}^{t-1} f^s_\tau
		\right]\inverse
		\PP_{x}
	},
\end{align}
a convenient general form for arbitrary $\phi$.

To evaluate the effective action \eqref{eq:second_order_action} and find the NLO imaginary offset we set $\phi=i\phi_1$ a spacetime constant.
Like before \eqref{eq:time independent f}, the auxiliary field terms become proprotional to the identity matrix and we can group terms into powers of
\begin{align}
	f^\pm = f^\pm_\tau = \exp\pm(\delta K + \delta \mu + \phi_1)
	\label{eq:time independent f1}
\end{align}
which has the nice property $(f^{\pm})\inverse = f^\mp$ so that we may treat the sign label as a true exponent.
Defining $\Delta t = t'-t$ we simplify to
\begin{align}
	\HH_{x't',xt} = \left.\partial_{x't'}\partial_{xt^{}}S[\phi]\right|_{\phi=i\phi_1}
	=
	\oneover{\tilde{U}} \delta_{x'x}\delta_{t't}
	& + \sum_s \tr{
		\left[
			\one + f^{-s N_t}
		\right]\inverse
		f^{-s \Delta t}
		\PP_{x'}
		f^{+s \Delta t}
		\left[
		\one + f^{+s N_t}
		\right]\inverse
		\PP_{x}
	}.
\end{align}
Since the hopping amplitudes are symmetric $K=K\transpose$, the matrices $f$ \eqref{eq:time independent f1} are too $f=f\transpose$.
Moreover, the projectors are symmetric $\PP=\PP\transpose$.
Because the sum is over $s\in\{\pm1\}$, the signs and inverses conspire so that the two traces are over a matrix and its transpose, and are therefore equal.
So, we can consolidate the traces and use the projectors to isolate needed matrix elements
\begin{align}
	\HH_{x't',xt}
	=
	\oneover{\tilde{U}} \delta_{x'x}\delta_{t't}
	& + 2 \tr{
		(\one+f^{-N_t})\inverse f^{-\Delta t} \PP_{x'}
		f^{+\Delta t}  (\one+f^{+N_t})\inverse
		\PP_x
	}
	\\
	=
	\oneover{\tilde{U}} \delta_{x'x}\delta_{t't}
	& + 2 
		\left[(\one+f^{-N_t})\inverse f^{-\Delta t}\right]_{xx'}
		\left[ f^{+\Delta t}  (\one+f^{+N_t})\inverse \right]_{x'x}.
\end{align}
We may quickly evaluate the matrix elements by a unitary transformation from the eigenbasis of $K$,
\begin{align}
	\left[ f^{\pm\Delta t}  (\one+f^{\pm N_t})\inverse \right]_{x'x}
	&=
	\sum_k \UU\adjoint_{x'k} \frac{e^{\pm \Delta t (\delta \epsilon_k + \phi_1 + \delta \mu)}}{1+e^{\pm \beta (\epsilon_k + \phi_1/\delta + \mu)}} \UU_{kx}
	&
	\UU\adjoint K \UU &= \epsilon_k.
\end{align}
We emphasize that these results rely on many simplifications offered by a constant spacetime offset $\phi = i \phi_1$.
However, nearly identical simplifications provide a similar evaluation for configurations with a different constant field on each temporal slice.

Rather than compute derivatives of the action expressed with $\log \det(\one+\FF_{\pm})$ \eqref{eq:schur} one may directly study $\log \det M_{\pm}$ instead.
In the case of constant field we may diagonalize $M$ with a straightforward unitary Matsubara decomposition
\begin{align}
	\LLambda_{kn,xt} &= \UU_{kx} e^{i \tilde{\omega}_n t} / \sqrt{N_t}
	&
	\tilde{\omega}_n &= (2n+1)\pi/N_t.
\end{align}
One finds a structurally similar and numerically equal expression for the Hessian in terms of matrix elements $T$
\begin{align}
	\HH_{x't',xt}
	&=
	\left( \frac{1}{\tilde U} - 1\right)\delta_{x',x}\delta_{t',t} - T_{+;xt,x't'}T_{+;x't',xt} - T_{-;xt,x't'}T_{-;x't',xt}
	\\
	T_{\pm;x't',xt} &= \sum_{kn} \LLambda\adjoint_{x't',kn} \frac{e^{\pm(\delta \epsilon_k + \delta \mu + \phi_1 + i \tilde{\omega}_n)}}{1-e^{\pm(\delta \epsilon_k + \delta \mu + \phi_1 + i \tilde{\omega}_n)}} \LLambda_{kn,xt}.
\end{align}
We have numerically verified that these two formulations yield the same Hessian.